\documentclass[showpacs,floatfix,superscriptaddress]{revtex4}
\usepackage{graphicx}
\usepackage{epsfig}
\usepackage{bm}
\usepackage[utf8]{inputenc}
\usepackage{amssymb}
\usepackage{float}
\usepackage{amsmath}
\usepackage{dcolumn}
\usepackage{float}
\usepackage{latexsym}
\usepackage{subfig}
\usepackage{tcolorbox}
\usepackage{amsthm}
\usepackage{booktabs}
\usepackage{rotating}
\usepackage{framed}
\usepackage{caption}
\usepackage[font=footnotesize,labelfont=bf]{caption}

\usepackage{cancel}
\usepackage{mathtools}

\usepackage[colorlinks]{hyperref}
\usepackage[dvipsnames]{xcolor}
\hypersetup{
     breaklinks=true,
    pdfstartview={FitH},    
    colorlinks=true,       
    linkcolor=blue,          
    citecolor=red,        
    filecolor=magenta,      
    urlcolor=blue,           
    anchorcolor=green,      
    linktocpage=true
}

\def\doi{http://doi.org}

\begin{document}

\title{Beyond $f(\phi)\mathcal{G}$: Gauss--Bonnet inflation with $\mu(\phi,X)$}

\author{Ali Seidabadi }
\email[]{aseidabadii@gmail.com}
\affiliation{Department of Theoretical Physics, Faculty of Sciences, University of Mazandaran,\\
47416-95447, Babolsar, Iran}

\author{Sara Saghafi}
\email[]{saghafisara1366@gmail.com}
\affiliation{Department of Theoretical Physics, Faculty of Sciences, University of Mazandaran,\\
47416-95447, Babolsar, Iran}
\affiliation{School of Physics, Damghan University, Damghan.\\
 3671645667, Iran}

\author{Kourosh Nozari}
\email[]{nozari7450@gmail.com (Corresponding Author)}
\affiliation{Department of Theoretical Physics, Faculty of Sciences, University of Mazandaran,\\
47416-95447, Babolsar, Iran}

\begin{abstract}
Gauss--Bonnet inflation typically affects the dynamics over an extended portion of the trajectory,
making it difficult to isolate a controlled imprint at CMB scales.
We consider a trajectory-selective coupling \(\mu(\phi,X)\) that gates the Gauss--Bonnet sector in phase space,
enabling the higher-curvature contribution to be localized within a finite e-fold window while remaining negligible elsewhere.
We identify stable inflationary solutions consistent with this localization and enforce standard ghost and gradient stability
conditions for both scalar and tensor perturbations.
For these viable backgrounds we compute pivot-scale observables and examine their dependence on the overall Gauss--Bonnet strength
and on the kinetic gating.
The framework offers a controlled route for realizing localized higher-curvature effects with predictable consequences for CMB-scale measurements.

\vspace{12 pt}

\end{abstract}

\maketitle

\enlargethispage{\baselineskip}
\tableofcontents

\section{Introduction}
\label{sec:introduction}

Inflation remains the most economical framework for generating the observed primordial perturbations, yet it is now
confronted with increasingly sharp empirical filters. Measurements of the scalar spectrum constrain departures from
scale invariance at the percent level, while primordial tensors are bounded to be small, pushing viable models toward
a relatively tight region in the $(n_s,r)$ plane. The Planck 2018 analysis provides stringent constraints on inflationary
tilt and running, and current BICEP/Keck likelihoods significantly restrict the tensor-to-scalar ratio at CMB scales.
More recently, the Atacama Cosmology Telescope (ACT) DR6 release has delivered high-precision temperature and
polarization spectra and competitive cosmological parameter constraints, strengthening the case that any new
high-energy ingredient during inflation must manifest as a controlled, correlated shift in multiple observables rather
than a simple rescaling of the scalar amplitude \cite{Planck2018Inflation,BKKeckXIII2021,ACTDR6LCDM2025}.

This observational situation sharpens a model-building challenge that is sometimes obscured by the flexibility of
single-field dynamics. Many ``beyond-GR'' effects can be hidden by retuning the potential or by mild changes in the
effective slow-roll history. If the goal is to isolate a distinctive signature of ultraviolet (UV) physics, one is naturally
led to higher-curvature operators whose origin is theoretically motivated and whose inflationary imprint is not
trivially degenerate with potential parameters. On the other hand,
in addition to the familiar possibility of a non-minimal coupling between the inflaton and the curvature sector,
which is well motivated for several compelling reasons~\cite{Faraoni2000}, one may also consider derivative
interactions that tie the inflaton kinematics directly to the background geometry~\cite{Amendola1993}.
At the level of the effective Lagrangian, this class is commonly represented by the operator
\(
\bigl[R_{\mu\nu}-\tfrac{1}{2}R g_{\mu\nu}\bigr]\nabla^{\mu}\phi\,\nabla^{\nu}\phi
\equiv G_{\mu\nu}\nabla^{\mu}\phi\,\nabla^{\nu}\phi,
\)
i.e. a coupling between the scalar derivatives and the Einstein tensor. It has been emphasized that in inflationary
setups with only a non-minimal scalar--curvature coupling, the background may probe energies above the unitarity
cutoff~\cite{BurgessLeeTrott2009,BarbonEspinosa2009,BurgessLeeTrott2010}. Subsequent work showed that
including the derivative coupling to the Einstein tensor can preserve (or effectively raise) the unitarity bound
during inflation~\cite{GermaniKehagias2010}.

From a complementary ultraviolet perspective, low-energy effective actions descending from superstring theory
generically contain higher-curvature corrections. Since the early Universe is a high-energy regime, such terms can
become dynamically relevant during inflation. In particular, the Gauss--Bonnet combination appears as the leading
curvature-squared correction in the stringy effective action~\cite{GrossSloan1987,CliftonFerreiraPadillaSkordis2012}.
Moreover, within string-inspired scalar--tensor constructions, the Gauss--Bonnet invariant naturally couples to
scalar fields, including an inflaton degree of freedom~\cite{NojiriOdintsovSasaki2005,SatohKannoSoda2008}.
The detailed cosmological impact then depends sensitively on the functional form of the scalar--Gauss--Bonnet
coupling, with pronounced consequences for both early- and late-time evolution~\cite{GuoSchwarz2010,Nojiri:2011,Nojiri:2017,OikonomouTsybaRazina2024,OikonomouEtAl2025}.

 Among curvature-squared corrections, the Gauss--Bonnet (GB)
invariant,
$\mathcal{G} \equiv R^2-4R_{\mu\nu}R^{\mu\nu}+R_{\mu\nu\rho\sigma}R^{\mu\nu\rho\sigma}$,
is particularly attractive. While it is topological in four dimensions when it appears alone, it becomes dynamically
relevant once coupled to a scalar degree of freedom, and it is well known to arise as an $\alpha'$ correction in
string-inspired effective actions. Moreover, the GB combination is distinguished by the fact that it avoids introducing
pathological higher-derivative degrees of freedom in the corresponding low-energy dynamics
\cite{Zwiebach1985,BoulwareDeser1985,GrossSloan1987}.

Inflation driven by a canonical scalar field with a GB coupling of the form $f(\phi)\,\mathcal{G}$ has therefore been widely
studied. In this setting, the GB sector modifies both the background evolution and the dynamics of the scalar and tensor
perturbations, allowing correlated shifts in $(n_s,r)$ and, in general, violation of the standard single-field consistency
relation. Early systematic treatments derived the primordial spectra and clarified the role of GB-induced corrections to
the tensor sector and the propagation speeds \cite{GuoSchwarz2009,GuoSchwarz2010,SatohSoda2008,Satoh2010GBCS}.
Recent analyses have revisited phenomenological viability using modern data combinations and updated pipelines,
highlighting parameter regions where the GB operator leaves an observable imprint while maintaining stability of the
fluctuations \cite{Kanti2015GBPlanck,Zhu2025GBData,Kawai:2023nqs,Koh2024HiggsGB}.
At the same time, renewed interest in transient amplifications of curvature perturbations (including regimes relevant
to enhanced small-scale power and primordial black hole scenarios) has motivated increasingly localized coupling
profiles and sharper diagnostics of when, along the inflationary trajectory, the GB operator effectively becomes active
\cite{KawaiKim2021PBHGB,AshrafzadehKarami2023,SolbiKarami2024_PTA_GB}.

A central limitation of the standard $f(\phi)\,\mathcal{G}$ setup, however, is that ``activation'' is controlled primarily by the field
space. Once the trajectory passes through a given $\phi$, the coupling is fixed irrespective of the instantaneous
phase-space state. This becomes conceptually restrictive when one wants to realize  \emph{trajectory-selective} effects,
for example localizing higher-curvature corrections to a finite e-fold window without contaminating the pivot-scale
dynamics elsewhere. More generally, short-lived departures from slow roll are intrinsically phase-space phenomena.
They depend not only on where the field is, but also on how it is moving. This motivates modifying the coupling to a
phase-space quantity, $\mu(\phi,X)$, with $X\equiv -\frac12 g^{\mu\nu}\partial_\mu\phi\partial_\nu\phi$, so that the GB
interaction can respond directly to the inflaton kinematics. Such kinetic sensitivity also connects naturally to the
broader scalar--tensor landscape in which derivative-dependent curvature couplings appear as consistent effective
operators within generalized scalar--tensor frameworks \cite{KobayashiYY2011,Kobayashi2019Review,EzquiagaKGB2016}.

Scalar--Gauss--Bonnet inflation has been confronted with post-Planck CMB data and studied in a variety of realizations
\cite{RashidiNozari2020AfterPlanck,Kawai:2021bye,NozariRashidi2017GBAlphaAttractor,JiangHuGuo2013InflationGB,KohLeeLeeTumurtushaa2014ObsGB}.
Reheating constraints in Gauss--Bonnet-coupled inflation have also been investigated
\cite{vandeBruckDimopoulosLongden2016ReheatingGB,BhattacharjeeMaityMukherjee2017ReheatingUnitarityPlanck,KohLeeTumurtushaa2018ReheatingGW}.
Related extensions combining Gauss--Bonnet effects with kinetic/derivative structures included
\cite{GrandaJimenezTorres2021KineticAndGB,PhamNguyenDo2021kGB},
and blue-tilted tensor spectra in derivative-driven super-inflation were discussed in
\cite{NozariShafizadeh2017BlueSpectrum}.

In this work we pursue this idea in a minimal and controlled way. We consider Einstein--scalar inflation with a
non-minimal GB interaction $\mu(\phi,X)\,\mathcal{G}$, and we design $\mu(\phi,X)$ to be \emph{phase-space gated}. The coupling is
localized in field space by a smooth bump profile while its effective strength is modulated by a bounded kinetic gate.
The practical aim is to separate two regimes within the same theoretical framework:
(i) a \emph{CMB-safe} regime where the pivot exits when the GB sector is effectively quiet and inflation is nearly
feature-free at CMB scales, and
(ii) a \emph{pivot-active} regime where higher-curvature physics leaves a visible imprint at horizon exit.
Crucially, we restrict attention to parameter regions where the explicit $X$-dependence acts as a \emph{perturbative}
controller near the pivot, so that the leading imprint of the gate enters through the modified background evolution
and the resulting time dependence of $\mu$ and its derivatives, while the standard quadratic-action expressions used in
scalar--GB inflation remain a reliable organizing principle \cite{GuoSchwarz2009,GuoSchwarz2010,SatohSoda2008}.
This controlled regime is then monitored a posteriori by enforcing ghost/gradient stability of both scalar and tensor
sectors along the entire trajectory and by checking that the kinetic modulation stays bounded.

Our phenomenological emphasis is deliberately pivot-centric. We adopt a fixed pivot prescription defined by a chosen
e-fold separation from the end of inflation and compute the pivot-scale observables from the scalar and tensor power
spectra at horizon exit. Rather than performing a broad, unconstrained scan over many parameters, we follow a
structured strategy that isolates the specific roles of the two physical ``knobs'' in the problem:
the overall GB strength and the kinetic gate. Concretely, we fix a representative coupling profile and vary the
overall GB strength by setting $\mu(\phi,X)\to \lambda_{\rm GB}\mu(\phi,X)$, and at fixed $\lambda_{\rm GB}$, scan the kinetic
gate amplitude controlling the explicit $X$-dependence. This enables a clean diagnosis of how $\lambda_{\rm GB}$ and the
phase-space controller shift $(n_s,r)$, and it clarifies when the GB sector influences the pivot primarily through its
overall strength versus through trajectory-selective modulation.

The paper is organized as follows. In Sec.~II we present the theoretical setup: the action, the background equations,
the slow-roll hierarchies, and the perturbation quantities and stability conditions used to evaluate the pivot-scale
spectra. In Sec.~III we specify the concrete phase-space gated coupling $\mu(\phi,X)$ by defining the bump profile and
the bounded kinetic gate, and we introduce the inflaton potential used in our numerical analysis. In Sec.~IV we
describe the numerical strategy, the pivot prescription, and the benchmark-based parameter scans. The resulting
background evolution, stability diagnostics, and pivot-scale predictions for both the $\lambda_{\rm GB}$ scan and the
kinetic-gate scan are presented in Sec.~V. We summarize the implications and outline natural extensions in Sec.~VI.

\section{Theoretical Setup}
\label{sec:model}

We consider a non-minimal Gauss--Bonnet inflationary model in four dimensions with a canonical scalar field
$\phi$ and a generalized Gauss--Bonnet coupling $\mu(\phi,X)$ that depends both on the field and its kinetic
term. Throughout we work in reduced Planck units $M_{\rm Pl}=1$ and metric signature $(-,+,+,+)$.

\subsection{Action with $\mu(\phi,X)$ Gauss--Bonnet Coupling}
\label{subsec:action}

In this setup, the action is given by 
\begin{equation}
S=\int d^4x \sqrt{-g}\left[
\frac{1}{2}R - \frac{1}{2}g^{\mu\nu}\partial_\mu\phi\,\partial_\nu\phi - V(\phi)
-\mu(\phi,X)\,\mathcal{G}
\right],
\label{eq:action}
\end{equation}
where $\mathcal{G}$ is the Gauss--Bonnet invariant,
\begin{equation}
\mathcal{G} = R^{\mu\nu\rho\sigma}R_{\mu\nu\rho\sigma}
- 4R^{\mu\nu}R_{\mu\nu}
+ R^2 .
\label{eq:GBinv}
\end{equation}
and the kinetic term is
\begin{equation}
X \equiv -\frac{1}{2}g^{\mu\nu}\partial_\mu\phi\,\partial_\nu\phi.
\label{eq:Xdef}
\end{equation}
Also, $R$ denotes the Ricci scalar, $\phi$ is the scalar field,  $\mu(\phi,X)$ describes the coupling function between the scalar field and Gauss--Bonnet curvature invariant .
\subsection{Background dynamics}
\label{sec:background}

Starting from the action \eqref{eq:action},
we adopt the spatially flat FLRW metric with a general lapse function as
\begin{align}
ds^2=-N^2(t)dt^2+a^2(t) d\vec{x}^{\,2}.
\label{eq:7}
\end{align}
The Hubble parameter is defined by
\begin{equation}
H\equiv \frac{1}{a}\frac{1}{N}\dot{a} = \frac{\dot{a}}{aN},
\label{eq:Hdef}
\end{equation}
where an overdot denotes differentiation with respect to the cosmic time $t$.
For a homogeneous scalar field $\phi(t)$ with canonical kinetic structure, we use
\begin{equation}
X \equiv -\frac{1}{2} g^{\mu\nu}\partial_\mu\phi\,\partial_\nu\phi= \frac{\dot{\phi}^{2}}{2N^{2}},
\label{eq:Xdef}
\end{equation}
the Ricci scalar is given by
\begin{align}
    R = 6 \left( \frac{\ddot{a}}{a N^{2}} 
- \frac{\dot{a}\, \dot{N}}{a N^{3}} 
+ \frac{\dot{a}^{2}}{a^{2} N^{2}} \right),
\end{align}
and the Gauss--Bonnet invariant takes the following form
\begin{align}
\mathcal{G} 
= 24\, \frac{\dot{a}^{2}}{a^{2} N^{2}}
\left( \frac{\ddot{a}}{a N^{2}} 
- \frac{\dot{a}\, \dot{N}}{a N^{3}} \right)
= 24 H^{2} \left( \frac{\dot{H}}{N} + H^{2} \right).
\label{eq:11}
\end{align}
To derive the background equations in a manifestly constrained form, we follow a minisuperspace (ADM) reduction.
Imposing FLRW symmetry at the level of the action, we substitute the metric \eqref{eq:7} into the
action \eqref{eq:action} and use $\sqrt{-g}=N a^{3}$ to obtain Lagrangian for the variables
$\{a(t),\phi(t),N(t)\}$ . It is convenient to decompose the resulting Lagrangian into the Einstein--Hilbert,
scalar-field, and Gauss--Bonnet contributions,

\begin{equation}
L = L_{\rm EH}+L_{\phi}+L_{\rm GB}
   = -\frac{3a\dot a^{2}}{N}+\frac{a^{3}\dot\phi^{2}}{2N}-N a^{3}V(\phi)
     -N a^{3}\mu(\phi,X)\,\mathcal{G}.
\label{eq:minisuperL0}
\end{equation}
Substituting the Gauss--Bonnet invariant given in \eqref{eq:11}
into the Lagrangian \eqref{eq:minisuperL0},
we obtain the final Lagrangian governing the background dynamics
\begin{align}
 L =    -\frac{3a\,\dot{a}^{2}}{N}
+ \frac{a^{3}}{2N}\,\dot{\phi}^{2}
- N a^{3} V(\phi)
-24\,a^{3}\,N\,\mu(\phi,X)\,
\frac{\dot a^{2}}{a^{2}N^{2}}
\left(
\frac{\ddot a}{aN^{2}}-\frac{\dot a \dot N}{aN^{3}}
\right)
\label{eq:Lfinal}
\end{align}

Starting from the Lagrangian \eqref{eq:Lfinal},
the background equations are obtained by variation with respect to
$N(t)$, $a(t)$, and $\phi(t)$.
After performing the variations, we fix the cosmic-time gauge $N=1$,
and the resulting equations can be written in the following forms:
\paragraph{Hamiltonian constraint (Friedmann equation)}
\begin{align}
    3H^{2}
= \frac{1}{2}\dot{\phi}^{2}
+ V(\phi)
- 24H^{3}\left( \mu_{,\phi}\,\dot{\phi}
+ \mu_{,X}\,\dot{\phi}\,\ddot{\phi} \right)
+ 48H^{4}\left( \frac{1}{2}\dot{\phi}^{2} \right)\mu_{,X},
\label{eq:14}
\end{align}
\paragraph{Acceleration equation (Raychaudhuri equation)}
\begin{align}
-2\dot H \;=\;& \dot\phi^{2}
+ 8H^{2}\left(\mu_{,\phi\phi}\dot\phi^{2}
              + \mu_{,\phi}\ddot\phi\right)
+ 16H\dot H\,\mu_{,\phi}\dot\phi
- 8H^{3}\mu_{,\phi}\dot\phi 
+ 16H^{2}\left(2\dot H - H^{2}\right) Y
+ 16H^{3}\dot Y.
\label{eq:15}
\end{align}
Here $\mu_{,\phi\phi}\equiv \partial^{2}\mu/\partial\phi^{2}$, $\mu_{,\phi}\equiv \partial\mu/\partial\phi$, $\mu_{,X}\equiv \partial\mu/\partial X$ and $Y \equiv X\mu_{,X}$ . 

\paragraph{Scalar-field equation}
\begin{align}
    (1 - \mu_X \mathcal{G})\, (\ddot{\phi} + 3H \dot{\phi})
\;-\;
\dot{(\mu_X \mathcal{G})}\, \dot{\phi}
\;+\;
V_{,\phi}
\;+\;
\mu_{,\phi} \mathcal{G}
= 0,
\label{eq:16}
\end{align}
where $V_{,\phi} \equiv \frac{\partial V}{\partial \phi}$ , 
$\dot{(\mu_X \mathcal{G})} = \dot{\mu}_X\,\mathcal{G} + \mu_X\, \dot{\mathcal{G}}$ and $\dot{\mathcal{G}}
= 24 \left[
    2H\,\dot{H}\,\left(\dot{H} + H^{2}\right)
    + H^{2}\left(\ddot{H} + 2H\,\dot{H}\right)
\right]$ . 
\\
\\
As consistency checks, we recover the following limiting cases.
\paragraph{Field-dependent Gauss--Bonnet coupling ($\mu=\mu(\phi)$).}
When the coupling depends only on the scalar field, $\mu(\phi,X)=\mu(\phi)$
(or equivalently $\mu_{,X}=0$), the background equations become
\begin{align}
&3H^{2} = \frac12 \dot\phi^{2} + V(\phi) - 24H^{3}\dot\mu(\phi), \\
&-2\dot H =
\dot\phi^{2}
+ 8H^{2}\ddot\mu(\phi)
+ 16H\dot H\,\dot\mu(\phi)
- 8H^{3}\dot\mu(\phi), \\
&\ddot\phi + 3H\dot\phi + V_{,\phi}
+ \mu_{,\phi}\,\mathcal{G} = 0 ,
\end{align}
where $\dot\mu(\phi)=\mu_{,\phi}\dot\phi$ and
$\ddot\mu(\phi)=\mu_{,\phi\phi}\dot\phi^{2}+\mu_{,\phi}\ddot\phi$.

\paragraph{Minimal single-field inflation ($\mu=0$).}
In the absence of the Gauss--Bonnet coupling, the background equations reduce to
\begin{align}
&3H^{2} = \frac12 \dot\phi^{2} + V(\phi), \\
&-2\dot H = \dot\phi^{2}, \\
&\ddot\phi + 3H\dot\phi + V_{,\phi} = 0 .
\end{align}

\subsection{Slow-roll approximation}

We now study the inflationary dynamics of the model in the slow-roll regime.
Starting from the background equations derived in the previous section,
we assume a quasi-de Sitter evolution in which the Hubble parameter and the
scalar field evolve slowly in time.
In the presence of a Gauss--Bonnet coupling $\mu(\phi,X)$, slow-roll inflation
is characterized not only by the smallness of the standard Hubble slow-roll
parameters, but also by the suppression of the time variation of the coupling
function and its phase-space gate.
We assume the usual inflationary slow-roll conditions
\begin{equation}
\dot{\phi}^{\,2} \ll V(\phi), \qquad |\ddot{\phi}| \ll 3H|\dot{\phi}|, \qquad
|\dot{H}| \ll H^2,
\end{equation}
and introduce the Hubble-flow hierarchy
\begin{equation}
\epsilon_1 \equiv -\frac{\dot{H}}{H^2}, \qquad
\epsilon_2 \equiv \frac{\dot{\epsilon}_1}{H\epsilon_1}.
\label{eps1}
\end{equation}
To quantify how strongly the GB sector contributes in the background, it is
useful to define four additional (dimensionless) hierarchies,
\begin{equation}
\delta_1 \equiv -8H\dot{\mu}, \qquad
\delta_2 \equiv \frac{\dot{\delta}_1}{H\delta_1}, \qquad
\kappa_1 \equiv 16H^2 Y = 16H^2 X\mu_{,X}, \qquad
\kappa_2 \equiv \frac{\dot{\kappa}_1}{H\kappa_1}.
\label{eq:deltakappa-def}
\end{equation}
In the slow-roll regime we take
\begin{equation}
|\epsilon_i|\ll 1, \qquad |\delta_i|\ll 1, \qquad |\kappa_i|\ll 1,
\end{equation}
so that products such as $\mathcal{O}(\delta_1\epsilon_1)$,
$\mathcal{O}(\delta_1\delta_2)$, and $\mathcal{O}(\kappa_1\kappa_2)$ can be
consistently neglected at leading order.
Under the slow-roll conditions, the background Eqs.(\ref{eq:14})--(\ref{eq:16}) reduce to
\begin{equation}
H^2 \simeq \frac{V(\phi)}{3}\left(1+\delta_1+\kappa_1\right)
\label{eq:SR-H2}
\end{equation}
\begin{align}
    -2\dot{H} \simeq \dot{\phi}^{2}+H^{2}(\delta_{1}-\kappa_{1}),
\end{align}
\begin{align}
    3H(1-\mu_{X}\mathcal{G})\dot{\phi} \simeq-V_{,\phi}-24\mu_{,\phi}H^{4}.
\end{align}
As consistency checks, in the case that $\mu=\mu(\phi)$, one recovers the well-known results

\begin{align}
    H^{2} \simeq \frac{V}{3},
\end{align}
\begin{align}
    -2\dot{H} \simeq \dot{\phi}^{2}-8H^{3}\mu_{,\phi}\dot{\phi},
\end{align}
\begin{align}
    3H\dot{\phi}+V_{,\phi} \simeq-24\mu_{,\phi}H^{4},
\end{align}
respectively.

\subsection{Linear perturbations, stability and pivot observables}
\label{sec:pert}

In our model the Gauss--Bonnet coupling is promoted to $\mu(\phi,X)$.
We restrict to the parameter regime in which the explicit $X$-dependence remains perturbative throughout the evolution,
so that it acts as a controlled modulation of the Gauss--Bonnet sector.
In this perturbative regime of the $X$-dependent modulation, we use the standard quadratic-action expressions of the
Gauss--Bonnet inflation ~\cite{GuoSchwarz2009,GuoSchwarz2010,SatohSoda2008,
AshrafzadehKarami2023,SolbiKarami2024_PTA_GB},
evaluated on the full background trajectory $\{\phi(t),H(t),\mu(\phi(t),X(t))\}$.
The rationale is that the leading imprint of the kinetic gate enters through the modified background
and the resulting profiles of $\dot\mu$ and $\ddot\mu$, while genuinely new operators in the quadratic action sourced
purely by the explicit $X$-dependence are higher order in the small modulation and are neglected.
This working assumption is monitored a posteriori by enforcing the stability conditions below.

Around a spatially flat FLRW background, the quadratic actions for the comoving curvature perturbation $\zeta$
and the transverse--traceless tensor modes $h_{ij}$ can be written as
\begin{align}
S^{(2)}_{s}
&=\int dt\, d^3x\, a^{3}\, Q_{s}
\left[\dot\zeta^{\,2}-\frac{c_{s}^{2}}{a^{2}}(\nabla\zeta)^{2}\right],
\label{eq:S2s}\\
S^{(2)}_{t}
&=\frac{1}{8}\int dt\, d^3x\, a^{3}\, Q_{t}
\left[\dot h_{ij}^{\,2}-\frac{c_{t}^{2}}{a^{2}}(\nabla h_{ij})^{2}\right].
\label{eq:S2t}
\end{align}
Here $Q_s$ and $Q_t$ are the scalar and tensor kinetic prefactors, while $c_s^2$ and $c_t^2$ are the corresponding propagation speeds of the perturbations.
The coefficients of the quadratic actions are written as (in reduced Planck units, $M_{\rm Pl}=1$)
\begin{align}
Q_s &= 12Q_t + 16\,\frac{\Sigma}{\Theta^{2}}\,Q_t^{\,2},
\qquad
c_s^{\,2} = \frac{1}{Q_s}\left[
\frac{16}{a}\frac{d}{dt}\!\left(\frac{aQ_t^{\,2}}{\Theta}\right)
-4c_t^{\,2}Q_t\right],
\label{eq:Qs_cs2_mu}\\[1mm]
Q_t &= \frac14\left(1-4H\dot\mu\right),
\qquad
c_t^{\,2} = \frac{1-4\ddot\mu}{1-4H\dot\mu},
\label{eq:Qt_ct_mu}
\end{align}
where
\begin{align}
\Sigma \equiv \frac12\dot\phi^{\,2}-3H^{2}+24H^{3}\dot\mu,
\qquad
\Theta \equiv H-6H^{2}\dot\mu .
\label{eq:Sigma_Theta_mu}
\end{align}

 The absence of ghosts and gradient instabilities requires
\begin{equation}
Q_s>0,\qquad c_s^2>0,\qquad Q_t>0,\qquad c_t^2>0.
\label{eq:stability}
\end{equation}

In Eqs.~\eqref{eq:Qt_ct_mu}--\eqref{eq:Sigma_Theta_mu}, $\dot\mu$ and $\ddot\mu$ denote total time derivatives.
For the phase-space gated coupling $\mu(\phi,X)$ with $X\equiv \dot\phi^{\,2}/2$, they are evaluated via the chain rule,
\begin{align}
\dot\mu &= \mu_{,\phi}\dot\phi+\mu_{,X}\dot X,
\qquad
\dot X=\dot\phi\,\ddot\phi,
\label{eq:muDot}\\
\ddot\mu &= \mu_{,\phi\phi}\dot\phi^{\,2}+\mu_{,\phi}\ddot\phi
+2\mu_{,\phi X}\dot\phi\,\dot X
+\mu_{,XX}\dot X^{2}+\mu_{,X}\ddot X.
\label{eq:muDDot}
\end{align}

The scalar and tensor power spectra at horizon exit are computed as
\begin{equation}
\mathcal P_s(k)=\left.\frac{H^2}{8\pi^2\,Q_s\,c_s^{\,3}}\right|_{c_s k=aH},
\qquad
\mathcal P_t(k)=\left.\frac{H^2}{2\pi^2\,Q_t\,c_t^{\,3}}\right|_{c_t k=aH},
\label{eq:PsPt}
\end{equation}
and the tensor-to-scalar ratio at the pivot is
\begin{equation}
r\equiv \frac{\mathcal P_t}{\mathcal P_s}
=4\,\frac{Q_s\,c_s^{\,3}}{Q_t\,c_t^{\,3}}.
\label{eq:rdef}
\end{equation}
The scalar spectral index is defined by
\begin{equation}
n_s-1\equiv \left.\frac{d\ln\mathcal P_s}{d\ln k}\right|_{c_s k=aH}.
\label{eq:nsdef}
\end{equation}
The pivot is specified by an e-fold separation $N_{\rm pivot}$ from the end of inflation,
\begin{equation}
N_\star \equiv N_{\rm end}-N_{\rm pivot},
\label{eq:Nstar}
\end{equation}
and all observables are evaluated at $N=N_\star$.
The corresponding comoving wavenumber is obtained a posteriori from the background mapping
\begin{equation}
k(N)=\frac{a(N)H(N)}{c_s(N)},
\qquad
\frac{d\ln k}{dN}=1+\frac{d\ln H}{dN}-\frac{d\ln c_s}{dN},
\label{eq:kofN}
\end{equation}
so that Eq.~\eqref{eq:nsdef} can be evaluated directly from a local derivative with respect to $N$,
\begin{equation}
n_s-1=\left.\frac{d\ln\mathcal P_s/dN}{\,d\ln k/dN\,}\right|_{N=N_\star}.
\label{eq:nsN}
\end{equation}

In addition to the pivot-scale estimates obtained from the slow-roll expressions, we also compute the primordial scalar and tensor spectra by directly evolving the linear perturbations on the same background solutions. This provides an explicit numerical cross-check of the perturbative treatment adopted in this work and clarifies the regime in which the slow-roll approximation remains reliable for the phase-space gated Gauss--Bonnet coupling. In the controlled regime considered here, the perturbation dynamics is organized by the same coefficients \(Q_s,c_s^2,Q_t,c_t^2\) introduced above, evaluated on the full \(\mu(\phi,X)\) background through the total derivatives \(\dot\mu\) and \(\ddot\mu\).

For the scalar sector, upon introducing the canonical Mukhanov--Sasaki variable  ~\cite{AshrafzadehKarami2023,SolbiKarami2024_PTA_GB}
\begin{equation}
v_k \equiv z_s\,\mathcal{R}_k,
\qquad
z_s \equiv a\sqrt{2Q_s c_s},
\qquad
d\tau_s \equiv \frac{c_s}{a}\,dt,
\label{eq:vkdef}
\end{equation}
the scalar mode equation takes the form
\begin{equation}
\frac{d^2 v_k}{d\tau_s^2}
+\left(
k^2-\frac{1}{z_s}\frac{d^2 z_s}{d\tau_s^2}
\right)v_k=0.
\label{eq:MSscalar}
\end{equation}
Similarly, for tensor perturbations, defining
\begin{equation}
u_k \equiv z_t\,h_k,
\qquad
z_t \equiv a\sqrt{2Q_t c_t},
\qquad
d\tau_t \equiv \frac{c_t}{a}\,dt,
\label{eq:ukdef}
\end{equation}
leads to
\begin{equation}
\frac{d^2 u_k}{d\tau_t^2}
+\left(
k^2-\frac{1}{z_t}\frac{d^2 z_t}{d\tau_t^2}
\right)u_k=0.
\label{eq:MStensor}
\end{equation}

The scalar and tensor power spectra can be obtained as
\begin{equation}
P_s(k) =\frac{k^3}{2\pi^2}\left|\mathcal{R}_k\right|^2,
\qquad
P_t(k)=4\,\frac{k^3}{2\pi^2}\left|h_k\right|^2,
\label{eq:MSspectra}
\end{equation}

from which the corresponding values of \(n_s\) and \(r\) are numerically extracted at the pivot scale, following Eqs.~\eqref{eq:rdef} and~\eqref{eq:nsdef}.

\section{The Model : Phase-Space Gated Gauss--Bonnet Coupling}
\label{sec:model}

We now specify the concrete functional form of the scalar--Gauss--Bonnet
coupling used throughout this work. Our aim is to localize the Gauss--Bonnet
feature in field space while allowing its effective strength to be modulated
along the phase-space trajectory through the kinetic term $X$.
Accordingly, we take the coupling function to be as
\begin{equation}
\mu(\phi,X)=e^{-\alpha_\mu \phi}\Bigl[1+A\,B\!\Bigl(\frac{\phi-\phi_\star}{\Delta}\Bigr)\Bigr]
\Bigl[1+g_{_{X}}\,G(X)\Bigr],
\label{eq:muModel}
\end{equation}
where $B$ is a smooth bump function that localizes the feature around
$\phi\simeq \phi_\star$ with width $\Delta$, while $G(X)$ is a bounded kinetic
gate controlling the activation of the coupling in phase space. Parameter $\alpha_\mu$ governs the smooth evolution of the coupling
away from the localized feature.
The parameter $A$ sets the amplitude of the localized feature introduced
by the bump function $B$, centered at $\phi=\phi_\star$ with width $\Delta$.
The dimensionless coefficient $g_{_{X}}$ controls the strength of the kinetic
gate $G(X)$ and determines the magnitude of the phase-space modulation.

\begin{figure}[h]
    \centering
    \includegraphics[width=0.6\linewidth]{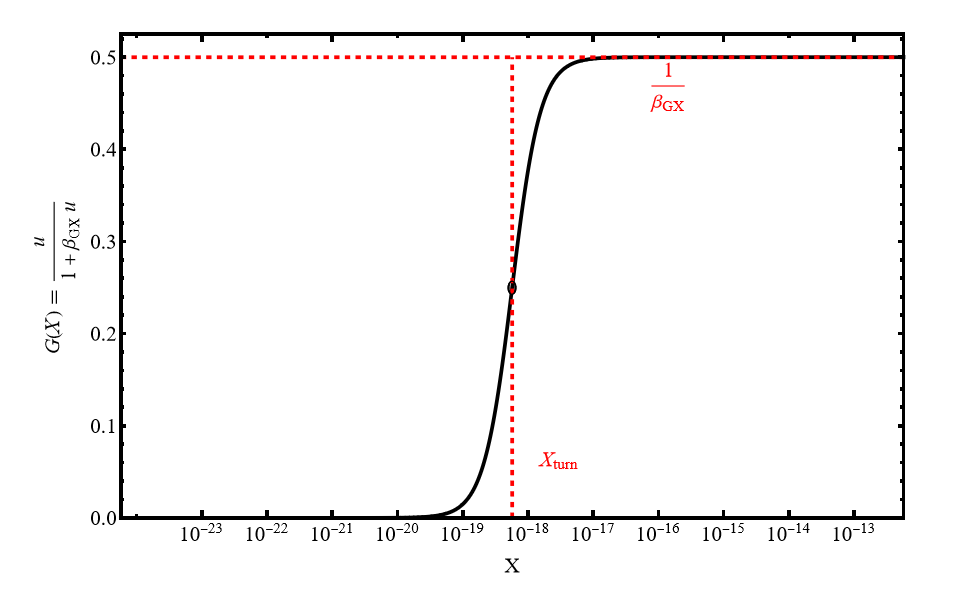}
    \caption{Kinetic gate function $G(X)=u/(1+\beta_{\rm GX}u)$ with $u=(X/M^{4})^{p}$.
The gate is bounded and monotonic, approaching the saturation level $G\to 1/\beta_{\rm GX}$
(red horizontal dashed line). The characteristic turn-on scale
$X_{\rm turn}\simeq M^{4}\beta_{\rm GX}^{-1/p}$ is indicated by the red vertical dashed line,
where $G(X_{\rm turn})=1/(2\beta_{\rm GX})$ (black marker).}
    \label{fig:ge1}
\end{figure}

We choose the bump profile as a $C^\infty$ compact-support function,
\begin{equation}
B(z)=
\begin{cases}
\exp\!\left[-\dfrac{1}{1-z^{2}}\right], & |z|<1,\\[6pt]
0, & |z|\ge 1,
\end{cases}
\qquad
z\equiv\frac{\phi-\phi_\star}{\Delta}.
\label{eq:bump}
\end{equation}
This choice ensures that the feature is smoothly switched on and off without
introducing spurious discontinuities in the background evolution.
For the kinetic gate we define
\begin{equation}
u(X)\equiv \left(\frac{X}{M^{4}}\right)^{p},
\qquad
G(X)\equiv \frac{u(X)}{1+\beta_{\rm GX}\,u(X)}.
\label{eq:gate}
\end{equation}
Here $u(X)$ is a dimensionless quantity constructed from the kinetic term
$X$ and the mass scale $M$. $p>0$ is a dimensionless parameter controlling how sharply the kinetic gate
turns on as a function of $X$ (larger $p$ corresponds to a steeper activation),
while $\beta_{\rm GX}>0$ is a dimensionless parameter that fixes the saturation
level of the gate, $G\to 1/\beta_{\rm GX}$ for $u\gg 1$.
The transition typically occurs around $\beta_{\rm GX}u(X)\sim 1$, i.e.
\begin{equation}
X_{\rm turn}\sim M^{4}\,\beta_{\rm GX}^{-1/p},
\end{equation}
so that $M$ sets the characteristic kinetic scale and $(p,\beta_{\rm GX})$
control the sharpness and the saturation of the phase-space modulation (see Fig.~\ref{fig:ge1}).

\begin{figure}[h]
    \centering
    \includegraphics[width=0.6\linewidth]{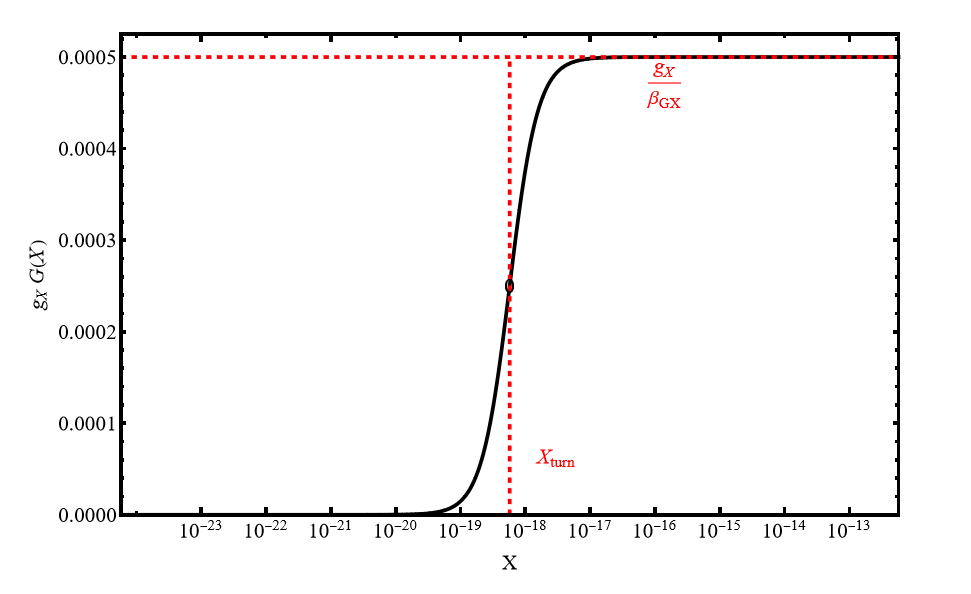}
    \caption{Saturation of the effective phase-space modulation $g_{_{X}}G(X)$.
The $X$-dependence is bounded, with $g_{_{X}}G(X)\to g_X/\beta_{\rm GX}$ for $X\gg X_{\rm turn}$
(red horizontal dashed line), while remaining suppressed for $X\ll X_{\rm turn}$.
The turn-on scale $X_{\rm turn}$ is marked by the red vertical dashed line.}
    \label{fig:ge2}
\end{figure}

The kinetic gate $G(X)=u/(1+\beta_{\rm GX}u)$ is bounded and monotonic,
with $G\to 0$ for $u\ll 1$ and $G\to 1/\beta_{\rm GX}$ for $u\gg 1$.
Accordingly, the combination $g_{_{X}} G(X)$ saturates to $g_{_{X}}/\beta_{\rm GX}$,
so the $X$-dependence provides a controlled phase-space modulation rather
than an unbounded deformation (see Fig.~\ref{fig:ge2}).

Having specified the form of the coupling function, we consider the following form of the scalar field potential \cite{KalloshLindeRoest2013} :
\begin{equation}
V(\phi)=V_0\left(1-e^{-b\phi}\right)^2,
\qquad
b=\sqrt{\frac{2}{3\alpha}} \, .
\label{eq:potential}
\end{equation}

Here $V_0$ sets the overall energy scale of inflation, while the parameter
$\alpha$ controls the curvature of the potential through
$b=\sqrt{2/(3\alpha)}$.

\section{Background evolution and numerical setup}
With the choices of the coupling function $\mu(\phi,X)$ and the scalar
potential $V(\phi)$ specified above, we proceed to study the background dynamics and perturbation properties
resulting from this setup.
For numerical evolution we work in $N\equiv\ln a$ and denote $d/dN$ by a prime.
The conversion rules are
\begin{align}
\frac{d}{dt}=H\frac{d}{dN},\qquad
\dot\phi=H\phi',\qquad
\ddot\phi=H^2\!\left(\phi''+\frac{H'}{H}\phi'\right),\qquad
\dot H=HH',
\end{align}
so that
\begin{align}
X=\frac{1}{2}H^2\phi'^2,\qquad
\mathcal{G}=24H^4\left(1+\frac{H'}{H}\right),\qquad
\dot\mu=H\mu',\qquad
\ddot\mu=H^2\mu''+HH'\mu',\qquad
\dot Y=HY'.
\end{align}

Substituting the coupling function \eqref{eq:muModel} and the potential
\eqref{eq:potential} into the background equations derived in
Sec.~\ref{sec:background}, we solve the system numerically using the
number of e-folds $N\equiv \ln a$ as the time variable. The initial conditions are chosen deep in the slow-roll regime, and the
integration is performed until the end of inflation, defined by
$\epsilon_1=1$. As specified in Sec.~\ref{sec:pert}, we adopt a fixed pivot prescription, and all observables reported are evaluated at the corresponding pivot time $N=N_\star$.
Since $V_0$ controls only the overall normalization of the inflationary energy scale,
we determine $V_0$ by imposing the observed scalar amplitude at the pivot scale
(COBE/Planck normalization). The parameter exploration is therefore performed over the
remaining dimensionless coupling parameters that control the shape and localized features of the model.
\\
To disentangle the dynamical impact of the phase-space gated Gauss--Bonnet sector from trivial parameter degeneracies,
we adopt a representative benchmark parameter set summarized in Table~\ref{tab:benchmark}.


\begin{table}[h]
\caption{Benchmark parameters used throughout the numerical analysis.
Unless stated otherwise, all figures are plotted using the same fixed parameter set below
(in reduced Planck units, $M_{\rm Pl}=1$).}
\label{tab:benchmark}
\centering
\setlength{\tabcolsep}{10pt} 
\renewcommand{\arraystretch}{1.15} 
\begin{tabular}{lcccc}
\hline\hline
Parameter & $\alpha_\mu$ & $\phi_\star$ & $\beta_{\rm GX}$ & $\alpha$ \\
\hline
Value     & $0.1$        & $5.3$        & $2$              & $1$      \\
\hline\hline
\end{tabular}
\end{table}

A key point is that we do \emph{not} search for arbitrary parameter sets.
Instead, we adopt a single benchmark (Table~\ref{tab:benchmark}) and perform two controlled scans:

\begin{enumerate}
\item \textbf{$\lambda_{\rm GB}$-scan (GB strength at fixed kinetic gate).}
We fix the kinetic-gate sector and vary the overall Gauss--Bonnet coupling strength by factoring the coupling as
$\lambda_{\rm GB}\,\mu(\phi,X)$, i.e.\ keeping the phase-space gating profile $\mu(\phi,X)$ fixed while scanning
$\lambda_{\rm GB}$ over a discrete set of values.
The bump-profile parameters $(\alpha_\mu, A, \phi_\star, \Delta)$ are held unchanged throughout this scan.
For each $\lambda_{\rm GB}$ we integrate the background equations up to the end of inflation, evaluate the
pivot-scale observables at $N=N_\star$, and verify the stability conditions along the trajectory.

\item \textbf{$g_{_{X}}$-scan (kinetic gate at fixed GB strength).}
We then fix the Gauss--Bonnet strength to a representative value,
$\lambda_{\rm GB}^{\rm fixed}=0.09$, and scan over the amplitude of the kinetic controller
at the pivot:
\begin{equation}
g_{_{X}}\in\{0,\,0.002,\,0.005,\,0.01,\,0.02,\,0.05,\,0.1\}.
\end{equation}

\end{enumerate}

In addition to the benchmark parameter set, we also consider the reference limits
$g_{_{X}}=0$ and $\lambda_{\rm GB}=0$. These baseline cases provide a direct point of comparison
for the figures and tables below, clarifying which features of the background evolution
and the pivot-scale quantities are genuinely induced by the Gauss--Bonnet sector and
the kinetic modulation.

\section{Numerical results}
\label{sec:results}

\begin{figure}[h]
    \centering
    \includegraphics[width=0.5\linewidth]{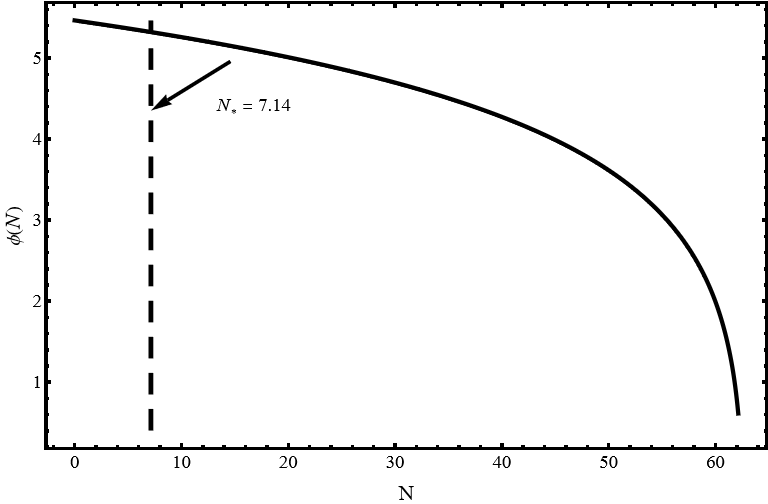}
    \caption{Background inflaton trajectory as a function of e-folds; $\phi(N)$, for a representative numerical solution of the phase-space gated Gauss--Bonnet model. The field rolls monotonically over the entire interval. The vertical dashed line marks the pivot scale $N_\star\simeq 7.14$ used for evaluating CMB-scale observables.}
    \label{phi}
\end{figure}

We now present representative numerical solutions of the phase-space gated Gauss--Bonnet model and discuss their physical implications at the pivot. Unless stated otherwise, the numerical evolution is performed using the benchmark set
$(\alpha_\mu,\phi_\star,\beta_{\rm GX},\alpha)$ in Table~\ref{tab:benchmark}, supplemented by the fixed choices
$A=0.05$, $\Delta = 0.9$, $p=2$, $M=3\times 10^{-5}$, $g_{_{X}}=10^{-3}$ and $\lambda_{GB}=10^{-3}$, together with the pivot choice
$N_\star = N_{\rm end}-N_{\rm pivot}$ with $N_{\rm pivot}=55$. the potential scale $V_0$ is determined by imposing the COBE/Planck normalization
$A_s(N_\star)=2.1\times10^{-9}$ at the pivot rather than treated as an independent scan parameter. Starting by displaying the background trajectory in terms of $N$, Fig.~\ref{phi} shows that the inflaton field starts from $\phi(N\simeq 0)\simeq 5.4$ and rolls down smoothly as $N$ increases, until inflation ends at $N_{\rm end}\simeq 62$.

\begin{figure}[h]
    \centering
    \includegraphics[width=0.5\linewidth]{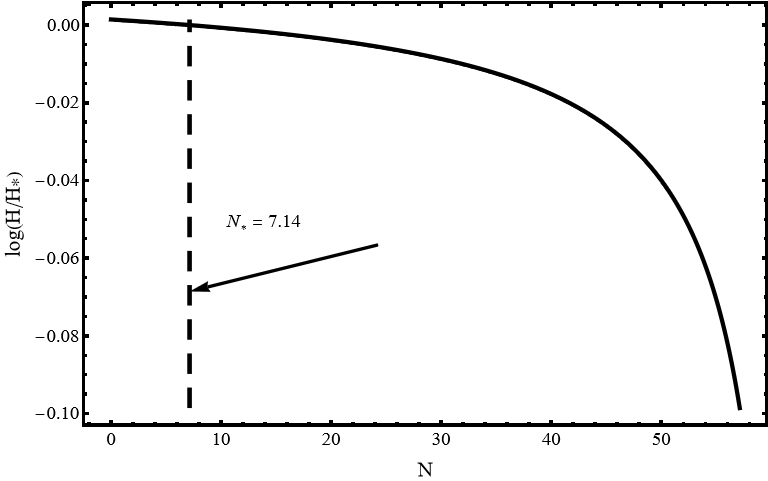}
    \caption{Hubble evolution along the background trajectory, shown in logarithmic scale as $\log(H/H_\star)$ with $H_\star\equiv H(N_\star)$. The vertical dashed line marks the pivot scale $N_\star\simeq 7.14$.}
    \label{fig:h}
\end{figure}

\begin{figure}[h]
    \centering
    \includegraphics[width=0.55\linewidth]{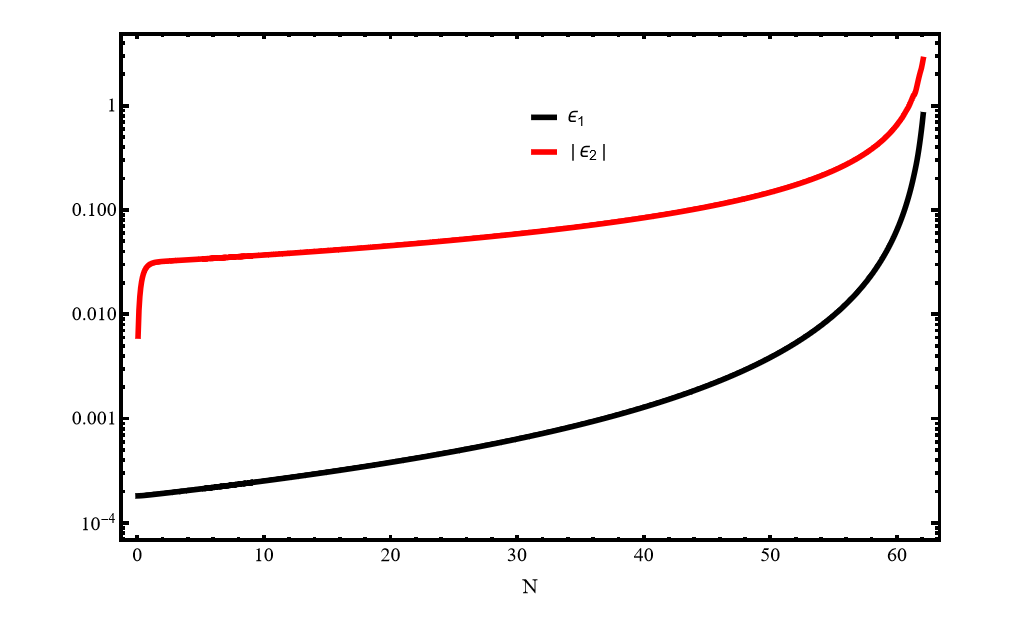}
    \caption{Evolution of the first two Hubble-flow (slow-roll) parameters, $\epsilon_1$ and $|\epsilon_2|$, along the background trajectory.}
    \label{fig:eps}
\end{figure}

Figure~\ref{fig:h} shows the Hubble evolution as $\log(H/H_\star)$, normalized to the pivot value $H_\star$. Around the pivot ($N_\star\simeq 7.14$) the curve remains very close to zero, indicating that $H$ stays nearly constant during the quasi--de Sitter stage. In this regime the expansion rate changes mildly. Toward the end of inflation the decline of $\log(H/H_\star)$ becomes noticeably steeper, signalling the departure from quasi--de Sitter evolution and the approach to the graceful exit where the first Hubble-flow parameter grows toward ${\cal O}(1)$. This transition is quantified by the Hubble-flow hierarchy. Figure~\ref{fig:eps} shows the first two slow-roll parameters, defined in Eq.~\eqref{eps1}, on a logarithmic scale. During the inflationary interval, both $\epsilon_1$ and $|\epsilon_2|$ remain below unity, consistent with the slow-roll evolution. In particular, $|\epsilon_2|$ varies smoothly, indicating that the rate at which $\epsilon_1$ evolves is not changing abruptly .

\begin{figure}[h]
\centering
\begin{minipage}[t]{0.48\linewidth}
  \centering
  \includegraphics[width=\linewidth]{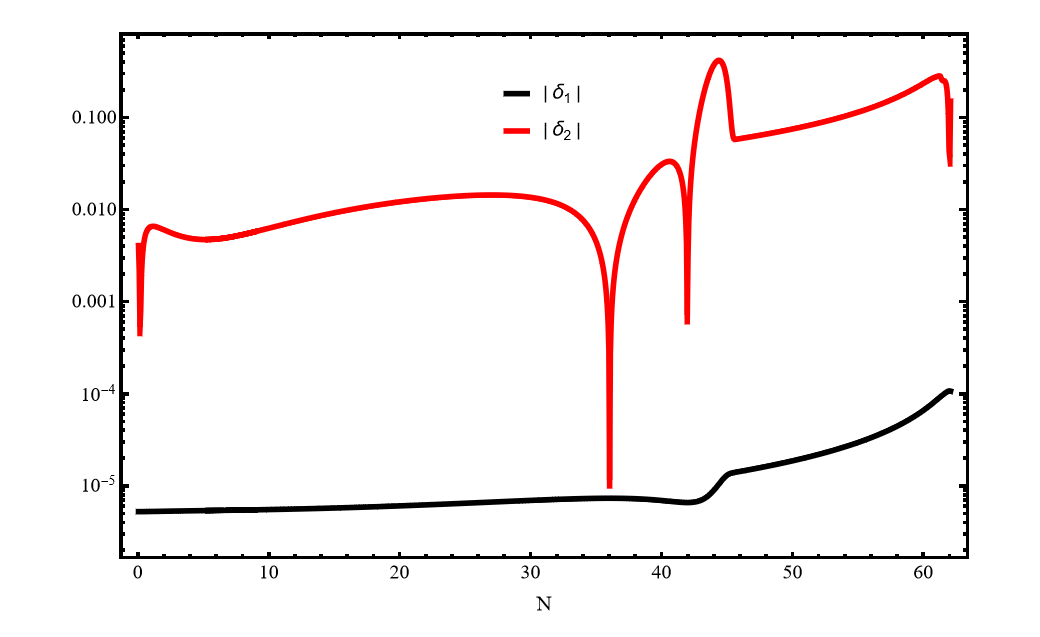}
  \captionof{figure}{Evolution of the Gauss--Bonnet slow-roll parameters $|\delta_1|$ and $|\delta_2|$ along the background trajectory.}
  \label{fig:delta12}
\end{minipage}\hfill
\begin{minipage}[t]{0.48\linewidth}
  \centering
  \includegraphics[width=\linewidth]{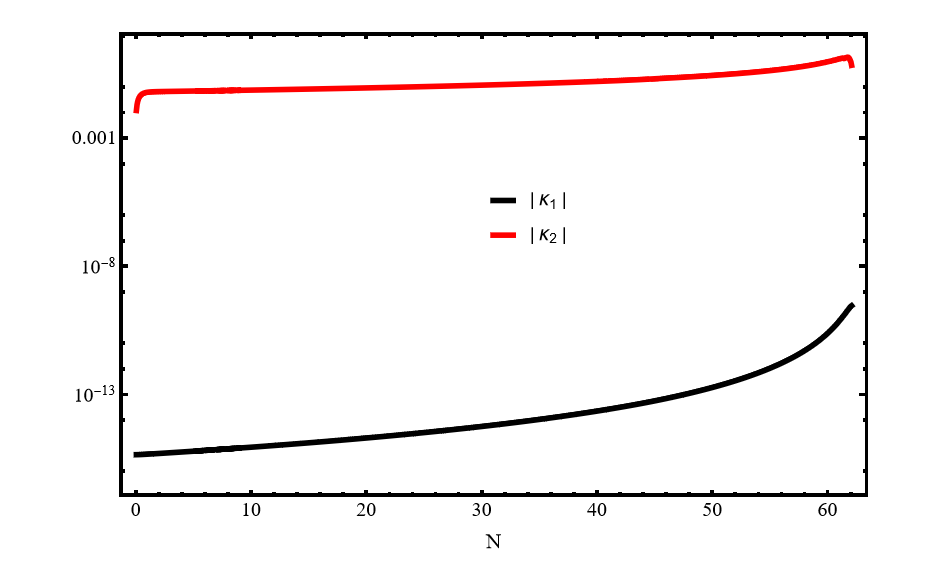}
  \captionof{figure}{Evolution of $|\kappa_1|$ and $|\kappa_2|$, which quantify the contribution of the explicit $X$-dependence of the Gauss--Bonnet coupling $\mu(\phi,X)$ along the background trajectory.}
  \label{fig:kappa12}
\end{minipage}
\end{figure}

\begin{figure}[h]
\centering
\begin{minipage}[t]{0.48\linewidth}
  \centering
  \includegraphics[width=\linewidth]{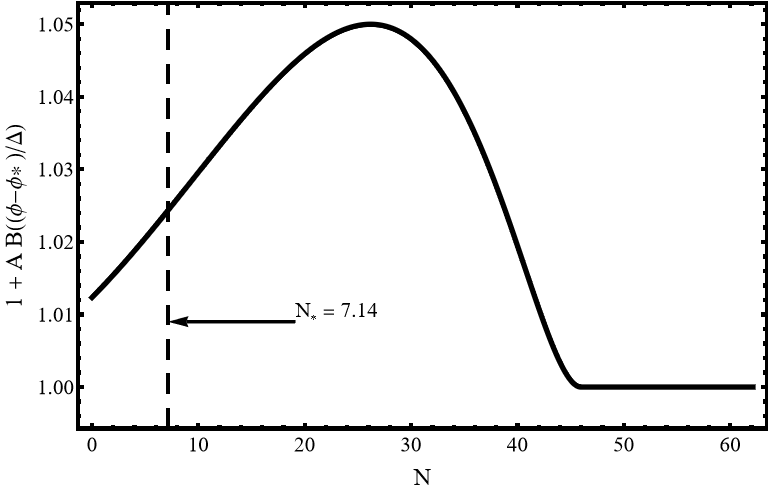}
  \captionof{figure}{Bump factor $1 + A\,B\!\big((\phi-\phi_\star)/\Delta\big)$ evaluated along the background trajectory as a function of $N$.}
  \label{fig:bump ba pivot}
\end{minipage}\hfill
\begin{minipage}[t]{0.515\linewidth}
  \centering
  \includegraphics[width=\linewidth]{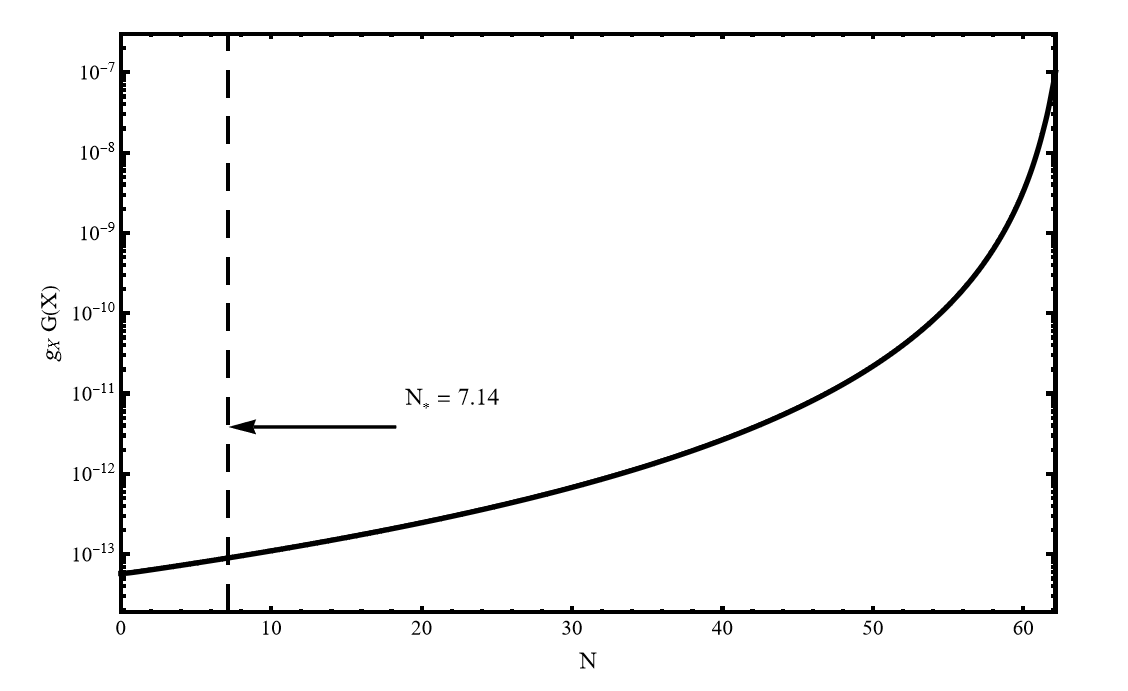}
  \captionof{figure}{Kinetic modulation $g_XG(X)$ evaluated along the background trajectory as a function of $N$.}
  \label{fig:g(x)}
\end{minipage}
\end{figure}

Including the Gauss--Bonnet sector, the slow-roll hierarchy can be supplemented by the additional parameters
defined in Eqs.~\eqref{eq:deltakappa-def}. Figures~\ref{fig:delta12} and \ref{fig:kappa12}  shows that
$|\delta_1|$, $|\delta_2|$, $|\kappa_1|$, and $|\kappa_2|$ also remain below unity over the interval shown,
indicating that the Gauss--Bonnet contribution remain small and under control in the background dynamics.
Moreover, the smallness of $(\kappa_1,\kappa_2)$ confirms that the explicit $X$-dependent part of
$\mu(\phi,X)$ stays perturbative along the trajectory. To make the phase-space gating explicit, Fig.~\ref{fig:bump ba pivot} shows the bump factor $1 + A\,B\!\big((\phi-\phi_\star)/\Delta\big)$ along the background trajectory. At the pivot $N_\star\simeq 7.14$ the factor remains close to unity, whereas its peak is well separated from the pivot scale. Accordingly, the Gauss--Bonnet enhancement is localized away from the pivot scale and acts only within a finite window in $N$, before the bump factor returns to unity. Furthermore, the kinetic modulation $g_{_{X}}G(X)$ is strongly suppressed around $N_\star$, reflecting the smallness of $X$ on pivot scale, so the $\phi$-dependent part of $\mu(\phi,X)$ dominates over the $X$-dependent contribution at the pivot. As $X(N)$ grows toward the end of inflation, $g_{_{X}}G(X)$ becomes larger, meaning that the kinetic gate is naturally weighted toward the exit stage rather than the pivot scale. Also , the kinetic modulation $g_{_{X}}G(X)$ remains well below unity, so this late-time enhancement still enters as a controlled correction. This late-time weighting is relevant for the post-inflationary transition, since the initial conditions for reheating are set by the background state at the end of inflation, where the kinetic energy is largest.

The Gauss--Bonnet contribution is more directly reflected in the coefficients of the quadratic actions for scalar and tensor perturbations, where even small departures from the GR limit can be quantified along the trajectory. In the GR limit ($\mu\to 0$) one has $Q_t\to M_{\rm pl}^2/4$ and $c_t^2\to 1$, while the scalar sector is governed by
$Q_s$ and $c_s^2$; departures from these limits therefore quantify how the phase-space gated coupling corrects the
perturbation dynamics. The evolution of $(Q_s,Q_t,c_s^2,c_t^2)$ indicates that the stability requirements are satisfied throughout the inflationary interval, while the Gauss--Bonnet corrections lead only to mild departures from the GR limit. In particular, $Q_s$ and $Q_t$ remain positive and the propagation speeds stay positive, with $c_s^2$ close to unity and the deviation of the tensor sector from its GR values remaining small along the trajectory. Furthermore, the brief dip in the propagation speeds can be understood as a transient imprint of the phase-space activation of the Gauss--Bonnet sector. In our setup, the bump in $\phi$ together with the kinetic gate in $X$ makes $\mu(\phi,X)$ rapidly time-dependent over a narrow e-fold interval, producing localized pulses in $\dot\mu$ and, in particular, $\ddot\mu$. Considering the tensor propagation speed squared given in Eq.~(\ref{eq:Qt_ct_mu}),
a short interval in which $\ddot\mu \gtrsim H\dot\mu$ naturally reduces $c_t^{\,2}$ below unity. The squared propagation speed of scalar perturbations inherits the same time dependence through the coefficients $Q_t$ and $\Theta$ defined in
Eqs.~(\ref{eq:Qt_ct_mu}) and~(\ref{eq:Sigma_Theta_mu})
so that the rapid variation of $Q_t^{2}/\Theta$ around the gate turn-on contributes a transient reduction in $c_s^{\,2}$. Once the system exits the feature (or the gate saturates), $\dot\mu,\ddot\mu\to 0$ and both $c_t^{\,2}$ and $c_s^{\,2}$ relax back toward their GR values.

\begin{figure}[h]
\centering
\begin{minipage}[t]{0.48\linewidth}
  \centering
  \includegraphics[width=\linewidth]{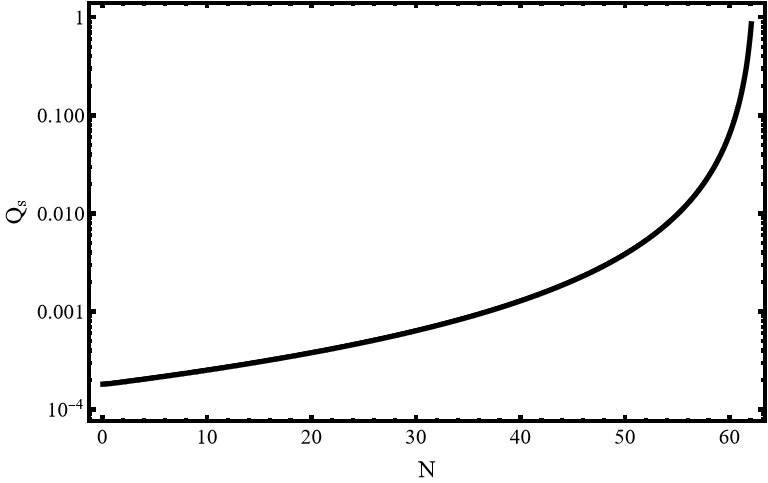}
  \captionof{figure}{Evolution of the scalar kinetic prefactor $Q_s(N)$ along the background trajectory.}
  \label{fig:Qs}
\end{minipage}\hfill
\begin{minipage}[t]{0.515\linewidth}
  \centering
  \includegraphics[width=\linewidth]{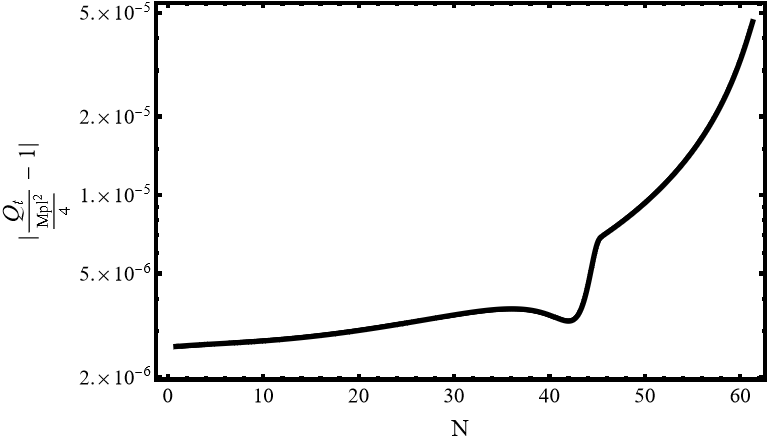}
  \captionof{figure}{Relative deviation of the tensor kinetic prefactor from its GR value, along the background trajectory.}
  \label{fig:Qt}
\end{minipage}
\end{figure}

\begin{figure}[h]
\centering
\begin{minipage}[t]{0.48\linewidth}
  \centering
  \includegraphics[width=\linewidth]{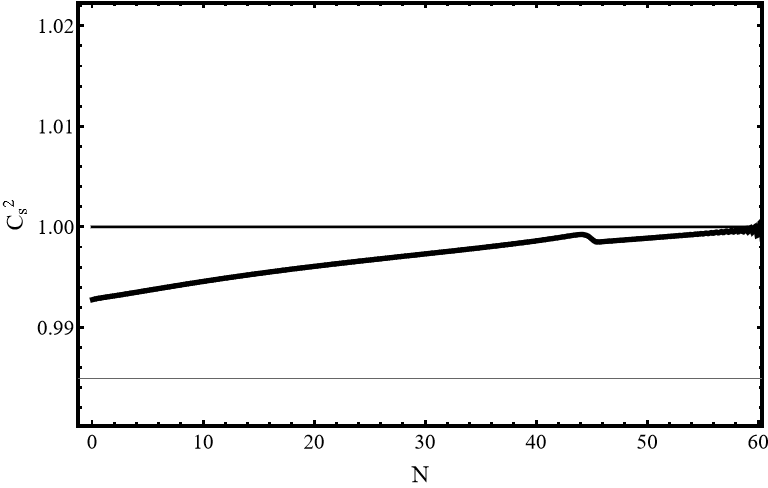}
  \captionof{figure}{Sound speed squared of scalar perturbations, $c_s^2(N)$, along the background trajectory.}
  \label{fig:cs}
\end{minipage}\hfill
\begin{minipage}[t]{0.515\linewidth}
  \centering
  \includegraphics[width=\linewidth]{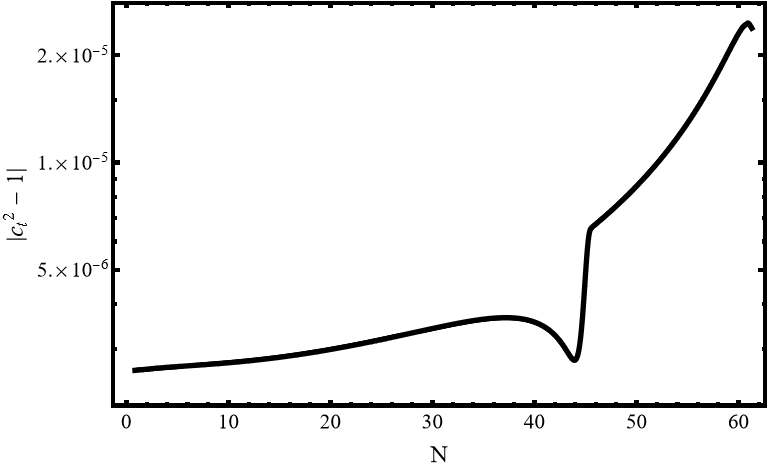}
  \captionof{figure}{Deviation of the tensor sound speed from unity, $|c_t^2(N)-1|$, along the background trajectory.}
  \label{fig:ct}
\end{minipage}
\end{figure}

These results show that Gauss--Bonnet effects enter the perturbation dynamics as controlled corrections without compromising stability. 
Their phenomenological impact is then assessed at the pivot by evaluating the pivot-scale observables $(n_s,r)$ from the scalar and tensor power spectra at $N_\star$, following Eqs.~\eqref{eq:rdef}--\eqref{eq:nsdef}; the corresponding results are summarized in Table~\ref{tab:pivot_lam_ns_r}.

\begin{table*}[h]
\caption{Pivot-scale observables for the $\lambda_{\rm GB}$ scan (values evaluated at $N_\star$).}
\label{tab:pivot_lam_ns_r}
\centering
\begin{tabular}{lcccc}
\hline\hline
$\lambda_{\rm GB}$ 
& $0$ 
& $5\times 10^{-4}$ 
& $10^{-3}$ 
& $2\times 10^{-3}$ \\
\hline
$n_s^{\rm slow\mbox{-}roll}$ 
& $0.964436$ 
& $0.964300$ 
& $0.964164$ 
& $0.963886$ \\
$n_s^{\rm MS}$ 
& $0.964882$ 
& $0.964749$ 
& $0.964610$ 
& $0.964336$ \\
$\Delta n_s$ 
& $0.000445388$ 
& $0.000448377$ 
& $0.000445444$ 
& $0.000450043$ \\
$r^{\rm slow\mbox{-}roll}$ 
& $0.00364300$ 
& $0.00362676$ 
& $0.00361066$ 
& $0.00357892$ \\
$r^{\rm MS}$ 
& $0.00354977$ 
& $0.00353393$ 
& $0.00351824$ 
& $0.00348729$ \\
$\Delta r$ 
& $-0.0000932254$ 
& $-0.0000928227$ 
& $-0.0000924198$ 
& $-0.0000916325$ \\
\hline\hline
\end{tabular}
\end{table*}

Table~\ref{tab:pivot_lam_ns_r} shows a mild but monotonic dependence of the pivot-scale observables on $\lambda_{\rm GB}$ at fixed kinetic-gate and bump-shape parameters. In the slow-roll evaluation, as $\lambda_{\rm GB}$ increases from $0$ to $2\times10^{-3}$, the scalar tilt decreases from $n_s^{\rm slow\mbox{-}roll}=0.964436$ to $n_s^{\rm slow\mbox{-}roll}=0.963886$, corresponding to a shift of about $-5.5\times10^{-4}$ and hence to a slightly redder spectrum at the pivot. Over the same range, the tensor-to-scalar ratio decreases from $r^{\rm slow\mbox{-}roll}=3.64300\times10^{-3}$ to $r^{\rm slow\mbox{-}roll}=3.57892\times10^{-3}$, indicating a small suppression of tensor power relative to the scalar sector.

The Mukhanov--Sasaki results in Table~\ref{tab:pivot_lam_ns_r} exhibit the same qualitative behavior. In particular, $n_s^{\rm MS}$ decreases monotonically from $0.964882$ to $0.964336$, while $r^{\rm MS}$ decreases from $3.54977\times10^{-3}$ to $3.48729\times10^{-3}$ across the same $\lambda_{\rm GB}$ interval. The corresponding differences between the two extractions remain small and smooth throughout the scan, with $\Delta n_s \sim 4.5\times10^{-4}$ and $\Delta r \sim -9\times10^{-5}$. Since the scalar amplitude is fixed at the pivot by construction, these shifts quantify the residual Gauss--Bonnet imprint on the pivot-scale observables within a regime where the background and perturbation stability conditions remain satisfied, while the direct numerical Mukhanov--Sasaki evolution confirms the same monotonic trend obtained from the slow-roll analysis. Furthermore, it is worth mentioning that typical CMB analyses report $n_s = 0.9649 \pm 0.0042$ (Planck 2018 TT,TE,EE+lowE)\cite{Planck2018Inflation}
and $\ln(10^{10}A_s)=3.044\pm 0.014$ at $k_\star=0.05\,{\rm Mpc}^{-1}$, with
$n_s=0.9665\pm0.0038$ when adding lensing+BAO\cite{Planck2018Inflation}; $B$-mode polarization further constrains tensors
(e.g. $r_{0.05}<0.036$ at 95\% C.L. from BICEP/Keck)\cite{BKKeckXIII2021}, while ACT DR6 reports
$n_s = 0.9743 \pm 0.0034$ and $\ln(10^{10}A_s)=3.053\pm0.013$ for the P-ACT-LB combination\cite{ACTDR6LCDM2025}.

Next, to make the Gauss--Bonnet imprint at the pivot more visible, the coupling parameters are updated to
$A=0.9$, $\Delta=0.45$, $p=12$, and $g_{_{X}}=0.03$, while all other parameters are kept fixed. This choice of the parameters ensures that the bump effect is localized around pivot scale $N_\star$ and clarifies the impact of the explicit $X$ dependence in the coupling. Considering these modified parameters, two scans are performed:
(i) a $\lambda_{\rm GB}$ scan at fixed $g_{_{X}}$, and (ii) a $g_{_{X}}$ scan at fixed $\lambda_{\rm GB}$.
For each scan point, the background evolution is integrated up to the end of inflation and the pivot-scale observables are evaluated at $N_\star$, while the pivot is changed to $N_{\rm pivot}=67$ .

\begin{figure}[h]
\centering
\begin{minipage}[t]{0.5\linewidth}
  \centering
  \includegraphics[width=\linewidth]{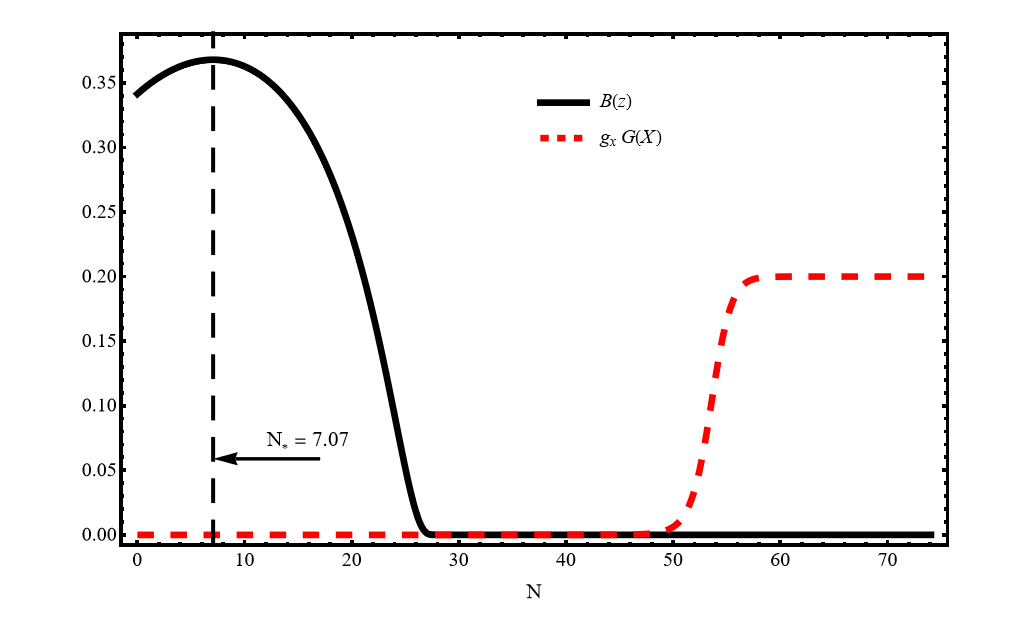}
  \captionof{figure}{Bump profile and kinetic modulation along the 
  background trajectory: $B(z)$ (bump sector) and $g_{_{X}}G(X)$ (kinetic gate). The vertical dashed line marks the pivot time $N_\star$.}
  \label{fig:gxbump}
\end{minipage}\hfill
\begin{minipage}[t]{0.5\linewidth}
  \centering
  \includegraphics[width=\linewidth]{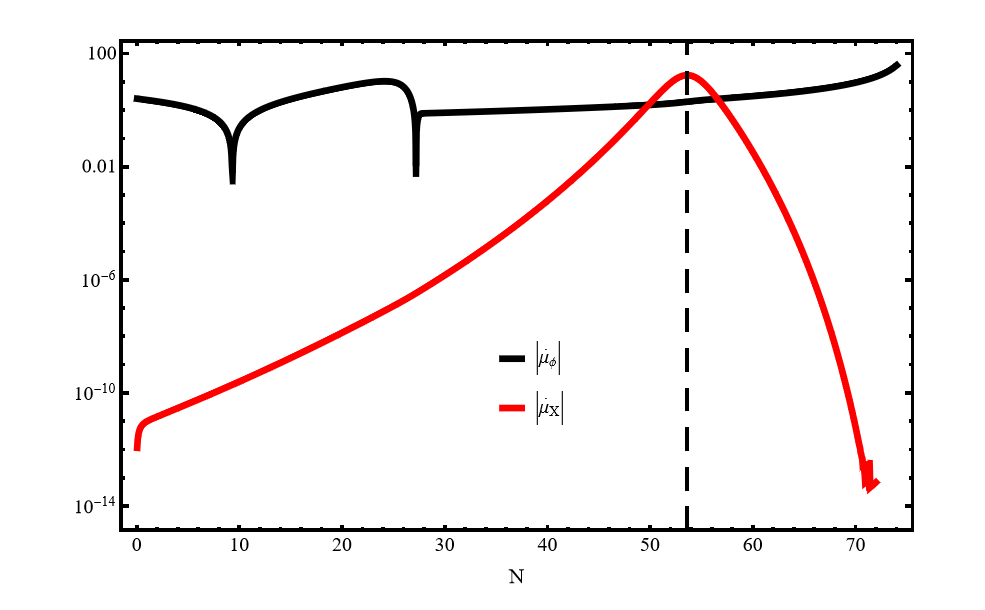}
  \captionof{figure}{Contributions $|\dot\mu_\phi|$ and $|\dot\mu_X|$ to the time variation of the coupling along the background trajectory. The vertical dashed line marks the late-time window where the $X$-driven contribution peaks.}
  \label{fig:mudotrelative}
\end{minipage}
\end{figure}

\begin{figure}[h]
\centering
\begin{minipage}[t]{0.5\linewidth}
  \centering
  \includegraphics[width=\linewidth]{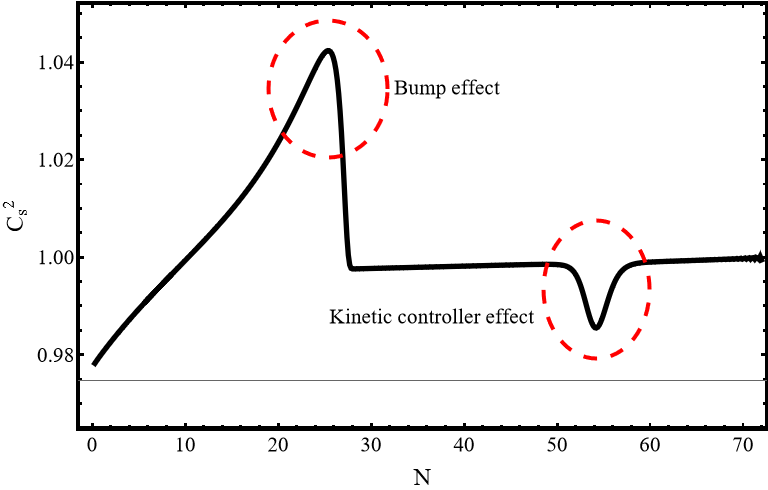}
  \captionof{figure}{Sound speed squared of scalar perturbations, $c_s^2(N)$, along the background trajectory. Two localized departures from $c_s^2=1$ occur in the bump and late-time controller windows.}
  \label{fig:csdayere}
\end{minipage}\hfill
\begin{minipage}[t]{0.5\linewidth}
  \centering
  \includegraphics[width=\linewidth]{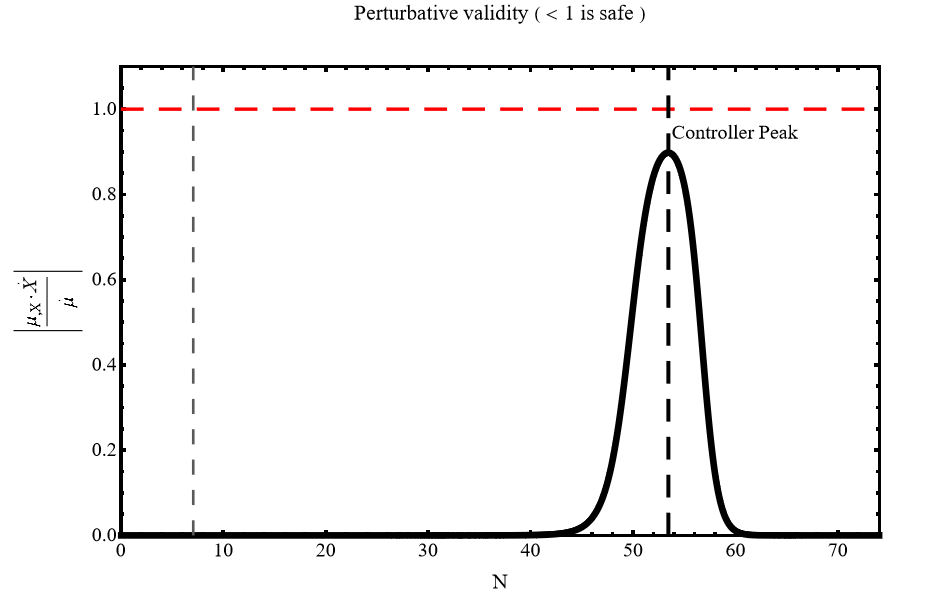}
  \captionof{figure}{Perturbative validity check for the $X$-dependent contribution, quantified by the ratio $|\mu_{,X}\dot X/\dot\mu|$ along the background trajectory. The vertical dashed line marks the pivot time $N_\star$ and the horizontal line indicates the reference perturbativity threshold.}
  \label{fig:validity}
\end{minipage}
\end{figure}

By fixing $\lambda_{GB} = 0.09$, as shown in Fig.~\ref{fig:gxbump}, the bump profile $B(z)$ and the kinetic modulation $g_{_{X}}G(X)$ are activated in well-separated e-fold windows along the background trajectory.
The bump contribution is confined to the pivot neighborhood around $N_\star$ and rapidly switches off afterward, whereas $g_{_{X}}G(X)$ is negligible at $N_\star$ and grows in relative importance only at later e-folds as the kinetic energy increases.
This separation in e-fold time isolates the pivot-localized bump effect from the late-time activation of the explicit $X$ dependence. In the same spirit, Fig.~\ref{fig:mudotrelative} compares the two contributions to the coupling time variation, $|\dot\mu_\phi|$ and $|\dot\mu_X|$, evaluated along the background trajectory.
Around the pivot one finds $|\dot\mu_\phi|\gg|\dot\mu_X|$, so the coupling evolution is effectively $\phi$-dominated on pivot scale, whereas the $X$-driven contribution becomes most relevant only in the late-time window where the kinetic gate is activated. Because the bump profile and the kinetic gate are activated in different e-fold intervals, one expects two distinct imprints on the scalar sound speed. Figure~\ref{fig:csdayere} follows this expectation. Around the bump window, $c_s^2$ increases above the GR value $c_s^2=1$, corresponding to a brief superluminal sound speed and indicating a Gauss--Bonnet-induced departure from the standard limit, whereas the late-time activation of the explicit $X$ dependence drives $c_s^2$ below unity in the controller window. Outside these localized intervals, $c_s^2$ remains positive and close to $1$. To make sure that the contribution of the $X$-dependent part remains perturbative, Fig.~\ref{fig:validity} verifies that $\mu_{,X}\dot X/\dot\mu \ll 1$ throughout the evolution.

\begin{table*}[h]
\caption{Pivot-scale observables for the $\lambda_{\rm GB}$ scan (values evaluated at $N_\star$).}
\label{tab:scanlambda}
\centering
\begin{tabular}{lcccccc}
\hline\hline
$\lambda_{\rm GB}$ 
& $0$ 
& $10^{-3}$ 
& $3\times 10^{-3}$ 
& $5\times 10^{-3}$ 
& $7\times 10^{-3}$ 
& $9\times 10^{-3}$ \\
\hline
$n_s^{\rm slow\mbox{-}roll}$ 
& $0.970724$ 
& $0.967836$ 
& $0.962261$ 
& $0.956824$ 
& $0.951460$ 
& $0.946165$ \\
$n_s^{\rm MS}$ 
& $0.971025$ 
& $0.968157$ 
& $0.962618$ 
& $0.957220$ 
& $0.951894$ 
& $0.946636$ \\
$\Delta n_s$ 
& $0.000301445$ 
& $0.000320879$ 
& $0.000356530$ 
& $0.000395049$ 
& $0.000433931$ 
& $0.000471493$ \\
$r^{\rm slow\mbox{-}roll}$ 
& $0.00248634$ 
& $0.00244342$ 
& $0.00236254$ 
& $0.00228740$ 
& $0.00221733$ 
& $0.00215197$ \\
$r^{\rm MS}$ 
& $0.00243387$ 
& $0.00239172$ 
& $0.00231231$ 
& $0.00223855$ 
& $0.00216975$ 
& $0.00210554$ \\
$\Delta r$ 
& $-0.0000524653$ 
& $-0.0000516961$ 
& $-0.0000502225$ 
& $-0.0000488466$ 
& $-0.0000475780$ 
& $-0.0000464371$ \\
\hline\hline
\end{tabular}
\end{table*}

\begin{table*}[h]
\caption{Pivot-scale observables for the $g_{_{X}}$ scan (values evaluated at $N_\star$).}
\label{tab:scang}
\centering
\begin{tabular}{lccccccc}
\hline\hline
$g_X$ 
& $0$ 
& $2\times 10^{-3}$ 
& $5\times 10^{-3}$ 
& $10^{-2}$ 
& $2\times 10^{-2}$ 
& $5\times 10^{-2}$ 
& $10^{-1}$ \\
\hline
$n_s^{\rm slow\mbox{-}roll}$ 
& $0.941268$ 
& $0.942908$ 
& $0.946548$ 
& $0.951205$ 
& $0.951773$ 
& $0.957058$ 
& $0.967679$ \\
$n_s^{\rm MS}$ 
& $0.941538$ 
& $0.943213$ 
& $0.946849$ 
& $0.951440$ 
& $0.952029$ 
& $0.957358$ 
& $0.967981$ \\
$\Delta n_s$ 
& $0.000269553$ 
& $0.000305833$ 
& $0.000300829$ 
& $0.000235091$ 
& $0.000256141$ 
& $0.000299399$ 
& $0.000301637$ \\
$r^{\rm slow\mbox{-}roll}$ 
& $0.00227074$ 
& $0.00224914$ 
& $0.00222961$ 
& $0.00223252$ 
& $0.00229032$ 
& $0.00238144$ 
& $0.00246229$ \\
$r^{\rm MS}$ 
& $0.00222122$ 
& $0.00220028$ 
& $0.00218152$ 
& $0.00218467$ 
& $0.00224112$ 
& $0.00233055$ 
& $0.00241018$ \\
$\Delta r$ 
& $-0.0000495106$ 
& $-0.0000488605$ 
& $-0.0000480869$ 
& $-0.0000478574$ 
& $-0.0000492018$ 
& $-0.0000508878$ 
& $-0.000052111$ \\
\hline\hline
\end{tabular}
\end{table*}

Consequently, by localizing the bump and kinetic modulation, the Gauss--Bonnet imprint becomes directly traceable in the pivot observables $(n_s,r)$ reported in Tables~\ref{tab:scanlambda} and \ref{tab:scang}. 
At fixed $g_X$, increasing $\lambda_{\rm GB}$ induces a clear red tilt. In Table~III, $n_s^{\rm slow\mbox{-}roll}$ decreases monotonically from $0.970724$ at $\lambda_{\rm GB}=0$ to $0.946165$ at $\lambda_{\rm GB}=9\times10^{-3}$, while the corresponding Mukhanov--Sasaki result follows the same trend, decreasing from $0.971025$ to $0.946636$. Over the same range, the tensor-to-scalar ratio is mildly suppressed, with $r^{\rm slow\mbox{-}roll}$ decreasing from $2.48634\times10^{-3}$ to $2.15197\times10^{-3}$ and $r^{\rm MS}$ from $2.43387\times10^{-3}$ to $2.10554\times10^{-3}$. The differences between the two extractions remain smooth and controlled across the scan, with $\Delta n_s$ increasing from $3.01445\times10^{-4}$ to $4.71493\times10^{-4}$, while $\Delta r$ stays at the level of a few $10^{-5}$. This confirms that the direct Mukhanov--Sasaki evolution reproduces the same red-tilt tendency and mild tensor suppression obtained from the slow-roll analysis.
By contrast, at fixed $\lambda_{\rm GB}=0.09$, increasing $g_X$ induces a monotonic shift toward a bluer tilt. As shown in Table \ref{tab:scang}, $n_s^{\rm slow\mbox{-}roll}$ rises from $0.941268$ at $g_X=0$ to $0.967679$ at $g_X=10^{-1}$, while $n_s^{\rm MS}$ rises in parallel from $0.941538$ to $0.967981$. Over the same scan, $r$ exhibits a milder but still monotonic increase, from $2.27074\times10^{-3}$ to $2.46229\times10^{-3}$ in the slow-roll treatment and from $2.22122\times10^{-3}$ to $2.41018\times10^{-3}$ in the Mukhanov--Sasaki calculation. The corresponding differences remain small, with $\Delta n_s$ staying of order $10^{-4}$ and $\Delta r$ remaining near $-5\times10^{-5}$ throughout the scan. Overall, the results indicate that $\lambda_{\rm GB}$ and the explicit $X$-dependence play distinct roles at the pivot: the overall Gauss--Bonnet strength drives the spectrum toward a redder tilt, whereas the kinetic controller shifts it back toward bluer values, while the full numerical Mukhanov--Sasaki treatment preserves the same qualitative trends.
Furthermore, to confirm that the reported results, and in particular their monotonic variation with $\lambda_{\rm GB}$ and $g_X$, remain valid under small variation of the benchmark parameters in Table~\ref{tab:benchmark}, we study a local sensitivity around the reference profile. In practice, we vary $(\alpha_\mu,\phi_\star,\beta_{GX},\alpha)$ and repeat the same numerical analysis for each parameter set. Tables~\ref{tab:q2_lambda_scan_combined} and ~\ref{tab:q2_gx_scan_combined} verifies that monotonic trends identified in the benchmark scans persist throughout the range of moderate variations considered here, while the corresponding pivot-scale observables undergo quantitative shifts. 
In the $\lambda_{\rm GB}$-scan of Table~\ref{tab:q2_lambda_scan_combined}, the monotonic shift toward a redder tilt of $n_s$ is preserved in all four representative cases. The $\alpha_\mu=0.5$ variation yields a milder shift toward a redder tilt, the $\beta_{GX}=4$ case remains close to the benchmark behavior, the displaced and narrower bump $(\phi_\star=5.8,\ \Delta=0.3)$ produces a substantially stronger shift toward a redder tilt, and the $\alpha=\frac{2}{3}$ case shifts the whole tensor sector downward while keeping the same monotonic tendency. In all these cases, $r$ continues to decrease monotonically with $\lambda_{\rm GB}$, and the Mukhanov--Sasaki values track the slow-roll trends with only quantitative offsets.
In the $g_X$-scan of Table ~\ref{tab:q2_gx_scan_combined}, $n_s$ again retains shift toward a bluer tilt as $g_X$ increases, but $r$ becomes more parameter-dependent. For $(\phi_\star=5.8,\ \Delta=0.4)$, the tensor-to-scalar ratio grows mildly with $g_X$; for $(\alpha=\frac{2}{3},\ \lambda_{\rm GB}=0.009)$, the shift toward a bluer tilt of $n_s$ becomes much steeper while $r$ decreases significantly; and for $\beta_{GX}=8$, the shift toward a bluer tilt of $n_s$ is maintained whereas the Mukhanov--Sasaki extraction yields a much stronger tensor suppression than the corresponding slow-roll values. 
\begin{table*}[h]
\caption{Pivot-scale observables for the $\lambda_{\rm GB}$ scan under mild variations of the benchmark parameters, comparing the slow-roll and Mukhanov--Sasaki results (values evaluated at $N_\star$).}
\label{tab:q2_lambda_scan_combined}
\centering
\begin{tabular}{lcccccc}
\hline\hline
$\lambda_{\rm GB}$
& $0$
& $10^{-3}$
& $3\times 10^{-3}$
& $5\times 10^{-3}$
& $7\times 10^{-3}$
& $9\times 10^{-3}$ \\
\hline
\multicolumn{7}{c}{$\alpha_\mu = 0.5$} \\
\hline
$n_s^{\rm slow\mbox{-}roll}$
& $0.970724$
& $0.970405$
& $0.969788$
& $0.969180$
& $0.968583$
& $0.967991$ \\
$n_s^{\rm MS}$
& $0.971025$
& $0.970715$
& $0.970101$
& $0.969496$
& $0.968899$
& $0.968314$ \\
$\Delta n_s$
& $0.000301445$
& $0.000310003$
& $0.000312733$
& $0.000316429$
& $0.000316178$
& $0.000323146$ \\
$r^{\rm slow\mbox{-}roll}$
& $0.00248634$
& $0.00247386$
& $0.00244941$
& $0.00242560$
& $0.00240242$
& $0.00237986$ \\
$r^{\rm MS}$
& $0.00243387$
& $0.00242164$
& $0.00239767$
& $0.00237434$
& $0.00235163$
& $0.00232952$ \\
$\Delta r$
& $-0.0000524653$
& $-0.0000522151$
& $-0.0000517330$
& $-0.0000512565$
& $-0.0000507956$
& $-0.0000503450$ \\
\hline
\multicolumn{7}{c}{$\beta_{GX} = 4$} \\
\hline
$n_s^{\rm slow\mbox{-}roll}$
& $0.970724$
& $0.967836$
& $0.962257$
& $0.956824$
& $0.951463$
& $0.946163$ \\
$n_s^{\rm MS}$
& $0.971025$
& $0.968155$
& $0.962617$
& $0.957220$
& $0.951894$
& $0.946636$ \\
$\Delta n_s$
& $0.000301445$
& $0.000318649$
& $0.000359297$
& $0.000395049$
& $0.000431389$
& $0.000473131$ \\
$r^{\rm slow\mbox{-}roll}$
& $0.00248634$
& $0.00244342$
& $0.00236254$
& $0.00228740$
& $0.00221734$
& $0.00215198$ \\
$r^{\rm MS}$
& $0.00243387$
& $0.00239172$
& $0.00231231$
& $0.00223855$
& $0.00216975$
& $0.00210554$ \\
$\Delta r$
& $-0.0000524653$
& $-0.0000516966$
& $-0.0000502248$
& $-0.0000488466$
& $-0.0000475825$
& $-0.0000464428$ \\
\hline
\multicolumn{7}{c}{$\phi_\star = 5.8,\ \Delta = 0.3$} \\
\hline
$n_s^{\rm slow\mbox{-}roll}$
& $0.970724$
& $0.963889$
& $0.950513$
& $0.937457$
& $0.924718$
& $0.912418$ \\
$n_s^{\rm MS}$
& $0.971025$
& $0.964209$
& $0.950869$
& $0.937817$
& $0.925025$
& $0.912580$ \\
$\Delta n_s$
& $0.000301445$
& $0.000320809$
& $0.000356047$
& $0.000359816$
& $0.000307816$
& $0.000161509$ \\
$r^{\rm slow\mbox{-}roll}$
& $0.00248634$
& $0.00243388$
& $0.00233657$
& $0.00224821$
& $0.00216800$
& $0.00209554$ \\
$r^{\rm MS}$
& $0.00243387$
& $0.00238223$
& $0.00228640$
& $0.00219903$
& $0.00211910$
& $0.00204617$ \\
$\Delta r$
& $-0.0000524653$
& $-0.0000516509$
& $-0.0000501676$
& $-0.0000491732$
& $-0.0000489046$
& $-0.0000493727$ \\
\hline
\multicolumn{7}{c}{$\alpha = \frac{2}{3}$} \\
\hline
$n_s^{\rm slow\mbox{-}roll}$
& $0.970502$
& $0.968490$
& $0.964568$
& $0.960790$
& $0.957190$
& $0.953795$ \\
$n_s^{\rm MS}$
& $0.970820$
& $0.968827$
& $0.964944$
& $0.961216$
& $0.957660$
& $0.954299$ \\
$\Delta n_s$
& $0.000318008$
& $0.000337362$
& $0.000376495$
& $0.000425277$
& $0.000469776$
& $0.000504239$ \\
$r^{\rm slow\mbox{-}roll}$
& $0.00170637$
& $0.00167882$
& $0.00162695$
& $0.00157924$
& $0.00153551$
& $0.00149561$ \\
$r^{\rm MS}$
& $0.00166998$
& $0.00164295$
& $0.00159207$
& $0.00154526$
& $0.00150238$
& $0.00146324$ \\
$\Delta r$
& $-0.0000363917$
& $-0.0000358705$
& $-0.0000348888$
& $-0.0000339747$
& $-0.0000331344$
& $-0.0000323707$ \\
\hline\hline
\end{tabular}
\end{table*}

\begin{table*}[h]
\caption{Pivot-scale observables for the $g_X$ scan under mild variations of the benchmark parameters, comparing the slow-roll and Mukhanov--Sasaki results (values evaluated at $N_\star$).}
\label{tab:q2_gx_scan_combined}
\centering
\begin{tabular}{lccccccc}
\hline\hline
$g_X$
& $0$
& $2\times 10^{-3}$
& $5\times 10^{-3}$
& $10^{-2}$
& $2\times 10^{-2}$
& $5\times 10^{-2}$
& $10^{-1}$ \\
\hline
\multicolumn{8}{c}{$\phi_\star = 5.8,\ \Delta = 0.4$} \\
\hline
$n_s^{\rm slow\mbox{-}roll}$
& $0.950171$
& $0.951612$
& $0.954095$
& $0.955844$
& $0.957630$
& $0.965806$
& $0.970724$ \\
$n_s^{\rm MS}$
& $0.950464$
& $0.951917$
& $0.954380$
& $0.956123$
& $0.957925$
& $0.966106$
& $0.971026$ \\
$\Delta n_s$
& $0.000292737$
& $0.000305470$
& $0.000285172$
& $0.000278939$
& $0.000294976$
& $0.000300765$
& $0.000301584$ \\
$r^{\rm slow\mbox{-}roll}$
& $0.00235276$
& $0.00234331$
& $0.00234142$
& $0.00236014$
& $0.00239604$
& $0.00245307$
& $0.00248634$ \\
$r^{\rm MS}$
& $0.00230213$
& $0.00229300$
& $0.00229132$
& $0.00230968$
& $0.00234488$
& $0.00240107$
& $0.00243387$ \\
\hline
\multicolumn{8}{c}{$\alpha = \frac{2}{3},\ \lambda_{\rm GB}=0.009$} \\
\hline
$n_s^{\rm slow\mbox{-}roll}$
& $0.950961$
& $0.951440$
& $0.952238$
& $0.953791$
& $0.957766$
& $0.977344$
& $1.03903$ \\
$n_s^{\rm MS}$
& $0.951269$
& $0.951800$
& $0.952651$
& $0.954300$
& $0.958460$
& $0.978707$
& $1.04069$ \\
$\Delta n_s$
& $0.000308051$
& $0.000360819$
& $0.000412853$
& $0.000508762$
& $0.000694502$
& $0.00136329$
& $0.00165260$ \\
$r^{\rm slow\mbox{-}roll}$
& $0.00159257$
& $0.00157284$
& $0.00154356$
& $0.00149561$
& $0.00140291$
& $0.00115118$
& $0.000820168$ \\
$r^{\rm MS}$
& $0.00155795$
& $0.00153868$
& $0.00151008$
& $0.00146324$
& $0.00137273$
& $0.00112752$
& $0.000808924$ \\
\hline
\multicolumn{8}{c}{$\beta_{GX} = 8$} \\
\hline
$n_s^{\rm slow\mbox{-}roll}$
& $0.941268$
& $0.942908$
& $0.946548$
& $0.951205$
& $0.951773$
& $0.957058$
& $0.960206$ \\
$n_s^{\rm MS}$
& $0.941538$
& $0.943213$
& $0.946846$
& $0.951440$
& $0.952029$
& $0.957358$
& $0.960505$ \\
$\Delta n_s$
& $0.000269553$
& $0.000305833$
& $0.000298230$
& $0.000235091$
& $0.000256141$
& $0.000300016$
& $0.000299796$ \\
$r^{\rm slow\mbox{-}roll}$
& $0.00227074$
& $0.00224914$
& $0.00222961$
& $0.00223252$
& $0.00229032$
& $0.00238144$
& $0.00240507$ \\
$r^{\rm MS}$
& $0.00111061$
& $0.00110014$
& $0.00109076$
& $0.00109233$
& $0.00112056$
& $0.00116528$
& $0.00117691$ \\
$\Delta r$
& $-0.00116012$
& $-0.00114900$
& $-0.00113885$
& $-0.00114019$
& $-0.00116976$
& $-0.00121616$
& $-0.00122815$ \\
\hline
\multicolumn{8}{c}{$\alpha_\mu = 0.5$} \\
\hline
$n_s^{\rm slow\mbox{-}roll}$
& $0.949506$
& $0.950051$
& $0.951072$
& $0.953282$
& $0.959291$
& $0.981584$
& $0.993600$ \\
$n_s^{\rm MS}$
& $0.949809$
& $0.950380$
& $0.951442$
& $0.953702$
& $0.959758$
& $0.981799$
& $0.993454$ \\
$\Delta n_s$
& $0.000303103$
& $0.000328643$
& $0.000369742$
& $0.000419427$
& $0.000467279$
& $0.000215675$
& $-0.000146280$ \\
$r^{\rm slow\mbox{-}roll}$
& $0.00197554$
& $0.00195874$
& $0.00193434$
& $0.00189603$
& $0.00182926$
& $0.00171710$
& $0.00176656$ \\
$r^{\rm MS}$
& $0.00193296$
& $0.00191656$
& $0.00189276$
& $0.00185543$
& $0.00179050$
& $0.00168202$
& $0.00173062$ \\
\hline\hline
\end{tabular}
\end{table*}

\section{Summary and Conclusions}\label{consum}

In this work we studied Gauss--Bonnet inflation where the coupling is modified from the standard
field-dependent form \(f(\phi)\,\mathcal{G}\) to a phase-space coupling function \(\mu(\phi,X)\,\mathcal{G}\), with
\(X=\dot\phi^2/2\).
The motivation is simple: once the coupling depends on \(X\) as well as \(\phi\), the Gauss--Bonnet sector no longer
tracks only field-space trajectory, but can respond to the instantaneous kinematics of the inflaton.
This provides a direct way to regulate when the higher-curvature operator becomes relevant along the background
trajectory.

The theoretical setup was developed at two complementary levels.
First, we derived and evolved the full background system implied by \(\mu(\phi,X)\), working in e-fold time \(N\) and
ending inflation at \(\epsilon_1=1\).
Second, we evaluated the perturbation quantities entering the scalar and tensor quadratic actions,
\((Q_s,c_s^2)\) and \((Q_t,c_t^2)\), using the standard Gauss--Bonnet expressions while inserting the full
\(\mu(\phi,X)\) background through the total derivatives \(\dot\mu\) and \(\ddot\mu\).
This strategy is appropriate in the regime where the explicit \(X\)-dependent modulation is perturbative, and we
validated this assumption by enforcing stability (\(Q_s>0\), \(c_s^2>0\), \(Q_t>0\), \(c_t^2>0\)) and by explicitly
checking that the \(X\)-driven contribution to \(\dot\mu\) remains subdominant when required.

To implement a controlled phase-space modulation, we adopted the bounded kinetic gate illustrated in
Figs.~\ref{fig:ge1} and~\ref{fig:ge2}. In particular, the monotonic activation around the characteristic scale
$X_{\rm turn}$ and the saturation $G\to 1/\beta_{\rm GX}$ ensure that the effective modulation $g_{_{X}}G(X)$ remains
finite (approaching $g_{_{X}}/\beta_{\rm GX}$ at large $X$), so the explicit $X$-dependence acts as a perturbative
controller rather than an unbounded deformation of the Gauss--Bonnet sector. We then fixed the benchmark
parameter set in Table~\ref{tab:benchmark} as a common reference profile for the bump/gate construction, and used
this baseline to perform the $\lambda_{\rm GB}$ and $g_X$ scans.

A central goal was to make the phase-space gating transparent in the numerical solutions.
In the benchmark evolution, the inflaton rolls monotonically and the expansion rate remains nearly constant during the
quasi--de Sitter stage, while the standard Hubble-flow parameters remain below
unity, confirming a controlled inflationary background (see Figs.~\ref{phi}--\ref{fig:h}--\ref{fig:eps}).
Including the Gauss--Bonnet sector, the additional hierarchies \((\delta_1,\delta_2)\) and \((\kappa_1,\kappa_2)\)
also remain below unity (see Figs.~\ref{fig:delta12} and \ref{fig:kappa12}), showing that the Gauss--Bonnet contribution to the background is small and that the explicit
\(X\)-dependence stays perturbative along the trajectory.
At the level of perturbations, \((Q_s,Q_t,c_s^2,c_t^2)\) stay in the stable domain while departing only mildly from
their GR limits (see Figs.~\ref{fig:Qs}--\ref{fig:cs}--\ref{fig:Qt}--\ref{fig:ct}), so the Gauss--Bonnet sector enters as a controlled correction rather than destabilizing the evolution.

The phenomenological imprint is most cleanly assessed at the pivot \(N_\star\), where we evaluate the observables using both the slow-roll expressions and the direct numerical Mukhanov--Sasaki evolution of the scalar and tensor modes.
For the first \(\lambda_{\rm GB}\) scan with fixed kinetic gate and bump-shape parameters, the results in
Table~\ref{tab:pivot_lam_ns_r} show a mild but monotonic shift as \(\lambda_{\rm GB}\) increases, corresponding to a
slightly redder scalar tilt together with a small suppression of \(r\).
This illustrates the expected behavior in a pivot-quiet regime, where the Gauss--Bonnet sector is present but remains
weakly active at horizon exit.

To make the Gauss--Bonnet imprint at the pivot more visible, we then updated the coupling parameters to
\(A=0.9\), \(\Delta=0.45\), \(p=12\), and \(g_{_{X}}=0.03\), while keeping the remaining benchmark parameters fixed and
evaluating the pivot observables at the revised pivot choice.
In this pivot-active configuration, the bump and the kinetic controller are activated in well-separated e-fold windows.
The bump is localized near \(N_\star\) while the kinetic modulation becomes relevant only at later times (see Figs.~\ref{fig:gxbump}), and the
relative contributions to \(\dot\mu\) confirm that the coupling time variation is \(\phi\)-dominated at the pivot and
\(X\)-driven only in the late-time controller window (see Figs.~\ref{fig:mudotrelative}).
As a result, the scalar sound speed exhibits two localized departures from the GR value \(c_s^2=1\), one associated
with the bump around the pivot and another correlated with the late-time controller (see Figs.~\ref{fig:csdayere}).
Throughout, a direct perturbativity check confirms that the \(X\)-dependent contribution remains a controlled
correction in the sense quantified in Fig.~\ref{fig:validity}.

The pivot observables in this regime show two clear and complementary behaviors.
At fixed \(g_{_{X}}\), increasing \(\lambda_{\rm GB}\) produces a monotonic redder tilt, with \(n_s\) decreasing across the
scan range reported in Table~\ref{tab:scanlambda}, while \(r\) decreases mildly over the same range.
At fixed \(\lambda_{\rm GB}\), increasing \(g_{_{X}}\) induces a monotonic shift toward a bluer tilt, with \(n_s\) rising
across the scan values in Table~\ref{tab:scang}, whereas \(r\) changes at a comparatively smaller level. To study local sensitivity, as summarized in Tables~\ref{tab:q2_lambda_scan_combined} and ~\ref{tab:q2_gx_scan_combined}, we considered mild variation of the benchmark parameter set and repeated the same numerical analysis. We found that the main monotonic trends of the benchmark $\lambda_{\rm GB}$ and $g_X$ scans remain intact, while the corresponding pivot-scale observables undergo quantitative shifts. This confirms that the reported benchmark behavior is locally robust within the controlled perturbative regime considered here.
Taken together, these scans show that \(\lambda_{\rm GB}\) and the explicit \(X\)-dependence play distinct roles at the
pivot. The overall Gauss--Bonnet strength primarily controls the tendency toward a redder spectrum, while the kinetic
controller provides an additional handle that can move the tilt back toward bluer values without spoiling stability.

There are several natural extensions.
First, the present analysis is restricted to the perturbative kinetic-modulation regime; pushing beyond it would require a
systematic reassessment of the quadratic action.
Second, since the kinetic controller becomes more important toward the end of inflation, it is well-motivated to
connect these phase-space effects to the post-inflationary transition and the subsequent reheating stage.
Overall, derivative-dependent Gauss--Bonnet couplings offer a concrete and controllable mechanism for making
higher-curvature physics phase-space selective, allowing one to assess a pivot-quiet regime or a pivot-active regime
while maintaining stability and producing a smooth, parameter-driven imprint on \((n_s,r)\).

\begin{acknowledgments}

A.S. is grateful to Milad Solbi for helpful discussions and insightful comments that helped improve this work. The authors are also grateful to the anonymous referee for valuable comments that helped improve the manuscript.

\end{acknowledgments}


\begin{thebibliography}{99}

\bibitem{Planck2018Inflation}
Y.~Akrami \textit{et al.} (Planck Collaboration),
Astron.\ Astrophys.\ \textbf{641}, A10 (2020)
[arXiv:1807.06211 [astro-ph.CO]]
\href{\doi/10.1051/0004-6361/201833887}{doi:10.1051/0004-6361/201833887}.

\bibitem{BKKeckXIII2021}
P.~A.~R.~Ade \textit{et al.} (BICEP/Keck Collaboration),
Phys.\ Rev.\ Lett.\ \textbf{127}, 151301 (2021)
[arXiv:2110.00483 [astro-ph.CO]]
\href{\doi/10.1103/PhysRevLett.127.151301}{doi:10.1103/PhysRevLett.127.151301}.

\bibitem{ACTDR6LCDM2025}
T.~Louis \textit{et al.} (The Atacama Cosmology Telescope Collaboration),
arXiv:2503.14452 [astro-ph.CO].

\bibitem{Faraoni2000}
V.~Faraoni,
Phys.\ Rev.\ D \textbf{62}, 023504 (2000).

\bibitem{Amendola1993}
L.~Amendola,
Phys.\ Lett.\ B \textbf{301}, 175--182 (1993).

\bibitem{BurgessLeeTrott2009}
C.~P.~Burgess, H.~M.~Lee and M.~Trott,
JHEP \textbf{09}, 103 (2009).

\bibitem{BarbonEspinosa2009}
J.~L.~F.~Barb\'on and J.~R.~Espinosa,
Phys.\ Rev.\ D \textbf{79}, 081302 (2009).

\bibitem{BurgessLeeTrott2010}
C.~P.~Burgess, H.~M.~Lee and M.~Trott,
JHEP \textbf{07}, 007 (2010).

\bibitem{GermaniKehagias2010}
C.~Germani and A.~Kehagias,
Phys.\ Rev.\ Lett.\ \textbf{105}, 011302 (2010).

\bibitem{CliftonFerreiraPadillaSkordis2012}
T.~Clifton, P.~G.~Ferreira, A.~Padilla, and C.~Skordis,
Phys.\ Rept.\ \textbf{513}, 1 (2012).

\bibitem{GrossSloan1987}
D.~J.~Gross and J.~H.~Sloan,
Nucl.\ Phys.\ B \textbf{291}, 41 (1987)
\href{\doi/10.1016/0550-3213(87)90465-2}{doi:10.1016/0550-3213(87)90465-2}.

\bibitem{NojiriOdintsovSasaki2005}
S.~Nojiri, S.~D.~Odintsov, and M.~Sasaki,
Phys.\ Rev.\ D \textbf{71}, 123509 (2005).

\bibitem{SatohKannoSoda2008}
M.~Satoh, S.~Kanno, and J.~Soda,
Phys.\ Rev.\ D \textbf{77}, 023526 (2008).

\bibitem{GuoSchwarz2010}
Z. -K.~Guo and D.~J.~Schwarz,
\href{\doi/10.1103/PhysRevD.81.123520}{Phys.\ Rev.\ D \textbf{81}, 123520 (2010)}
[arXiv:1001.1897 [hep-th]].

\bibitem{Nojiri:2011}
S.~Nojiri and S.~D.~Odintsov,
``Unified cosmic history in modified gravity: from F(R) theory to Lorentz non-invariant models,''
Phys.\ Rept.\ \textbf{505}, 59--144 (2011)
\href{https://doi.org/10.1016/j.physrep.2011.04.001}{doi:10.1016/j.physrep.2011.04.001}
[arXiv:1011.0544 [gr-qc]].

\bibitem{Nojiri:2017}
S.~Nojiri, S.~D.~Odintsov and V.~K.~Oikonomou,
``Modified Gravity Theories on a Nutshell: Inflation, Bounce and Late-time Evolution,''
Phys.\ Rept.\ \textbf{692}, 1--104 (2017)
\href{https://doi.org/10.1016/j.physrep.2017.06.001}{doi:10.1016/j.physrep.2017.06.001}
[arXiv:1705.11098 [gr-qc]].

\bibitem{OikonomouTsybaRazina2024}
V.~K.~Oikonomou, P.~Tsyba, and O.~Razina,
Annals Phys.\ \textbf{462}, 169597 (2024).

\bibitem{OikonomouEtAl2025}
V.~K.~Oikonomou, A.~Gkioni, I.~Sdranis, P.~Tsyba, and O.~Razina,
Class.\ Quant.\ Grav.\ \textbf{42}, 075016 (2025).

\bibitem{Zwiebach1985}
B.~Zwiebach,
Phys.\ Lett.\ B \textbf{156}, 315 (1985)
\href{\doi/10.1016/0370-2693(85)91616-8}{doi:10.1016/0370-2693(85)91616-8}.

\bibitem{BoulwareDeser1985}
D.~G.~Boulware and S.~Deser,
Phys.\ Rev.\ Lett.\ \textbf{55}, 2656 (1985)
\href{\doi/10.1103/PhysRevLett.55.2656}{doi:10.1103/PhysRevLett.55.2656}.

\bibitem{Satoh2010GBCS}
M.~Satoh,
Phys.\ Rev.\ D \textbf{81}, 023511 (2010)
[arXiv:0911.5191 [astro-ph.CO]]
\href{\doi/10.1103/PhysRevD.81.023511}{doi:10.1103/PhysRevD.81.023511}.

\bibitem{GuoSchwarz2009}
Z. -K.~Guo and D.~J.~Schwarz,
\href{\doi/10.1103/PhysRevD.80.063523}{Phys.\ Rev.\ D \textbf{80}, 063523 (2009)}
[arXiv:0907.0427 [hep-th]].

\bibitem{SatohSoda2008}
M.~Satoh and J.~Soda,
\href{\doi/10.1088/1475-7516/2008/09/019}{JCAP \textbf{09}, 019 (2008)}
[arXiv:0806.4594 [astro-ph]].

\bibitem{Zhu2025GBData}
Y.~Zhu, Q.~Gao, Y.~Gong, and Z.~Yi,
Eur.\ Phys.\ J.\ C \textbf{85}, 1227 (2025)
[arXiv:2508.09707 [astro-ph.CO]]
\href{https://doi.org/10.1140/epjc/s10052-025-14969-2}{doi:10.1140/epjc/s10052-025-14969-2}.

\bibitem{Kawai:2023nqs}
S.~Kawai and J.~Kim,
``Probing the inflationary moduli space with gravitational waves,''
Phys.\ Rev.\ D \textbf{108}, no.~10, 103537 (2023)
\href{https://doi.org/10.1103/PhysRevD.108.103537}{doi:10.1103/PhysRevD.108.103537}
[arXiv:2308.13272 [astro-ph.CO]].


\bibitem{Kanti2015GBPlanck}
P.~Kanti, R.~Gannouji, and N.~Dadhich,
Phys.\ Rev.\ D \textbf{92}, 041302 (2015)
[arXiv:1503.01579 [hep-th]]
\href{\doi/10.1103/PhysRevD.92.041302}{doi:10.1103/PhysRevD.92.041302}.

\bibitem{Koh2024HiggsGB}
S.~Koh, S.~C.~Park, and G.~Tumurtushaa,
Phys.\ Rev.\ D \textbf{110}, 023523 (2024)
[arXiv:2308.00897 [gr-qc]]
\href{https://doi.org/10.1103/PhysRevD.110.023523}{doi:10.1103/PhysRevD.110.023523}.

\bibitem{KawaiKim2021PBHGB}
S.~Kawai, J.~Kim,
Phys.\ Rev.\ D \textbf{104}, 083545 (2021)
[arXiv:2108.01340 [astro-ph.CO]]
\href{\doi/10.1103/PhysRevD.104.083545}{doi:10.1103/PhysRevD.104.083545}.

\bibitem{AshrafzadehKarami2023}
A.~Ashrafzadeh and K.~Karami,
Astrophys.\ J.\ \textbf{965}, 11 (2024)
[arXiv:2309.16356 [astro-ph.CO]]
\href{\doi/10.3847/1538-4357/ad293f}{doi:10.3847/1538-4357/ad293f}.

\bibitem{SolbiKarami2024_PTA_GB}
M.~Solbi and K.~Karami,
\href{\doi/10.1140/epjc/s10052-024-13271-x}{Eur.\ Phys.\ J.\ C \textbf{84}, 918 (2024)}
[arXiv:2403.00021 [gr-qc]].

\bibitem{KobayashiYY2011}
T.~Kobayashi, M.~Yamaguchi, and J.~Yokoyama,
Prog.\ Theor.\ Phys.\ \textbf{126}, 511 (2011)
[arXiv:1105.5723 [hep-th]]
\href{\doi/10.1143/PTP.126.511}{doi:10.1143/PTP.126.511}.

\bibitem{Kobayashi2019Review}
T.~Kobayashi,
Rept.\ Prog.\ Phys.\ \textbf{82}, 086901 (2019)
[arXiv:1901.07183 [gr-qc]]
\href{\doi/10.1088/1361-6633/ab2429}{doi:10.1088/1361-6633/ab2429}.

\bibitem{EzquiagaKGB2016}
J.~M.~Ezquiaga, J.~Garc{\'i}a-Bellido, and M.~Zumalac{\'a}rregui,
Phys.\ Rev.\ D \textbf{94}, 024005 (2016)
[arXiv:1603.01269 [gr-qc]]
\href{\doi/10.1103/PhysRevD.94.024005}{doi:10.1103/PhysRevD.94.024005}.

\bibitem{RashidiNozari2020AfterPlanck}
N.~Rashidi and K.~Nozari,
Astrophys.\ J.\ \textbf{890}, 58 (2020),
arXiv:2001.07012 [astro-ph.CO].

\bibitem{Kawai:2021bye}
S.~Kawai and J.~Kim,
``CMB from a Gauss-Bonnet-induced de Sitter fixed point,''
Phys.\ Rev.\ D \textbf{104}, no.~4, 043525 (2021)
\href{https://doi.org/10.1103/PhysRevD.104.043525}{doi:10.1103/PhysRevD.104.043525}
[arXiv:2105.04386 [hep-ph]].

\bibitem{NozariRashidi2017GBAlphaAttractor}
K.~Nozari and N.~Rashidi,
Phys.\ Rev.\ D \textbf{95}, 123518 (2017),
arXiv:1705.02617 [astro-ph.CO].

\bibitem{JiangHuGuo2013InflationGB}
P. -X.~Jiang, J. -W.~Hu, and Z. -K~Guo,
Phys.\ Rev.\ D \textbf{88}, 123508 (2013),
arXiv:1310.5579 [hep-th].

\bibitem{KohLeeLeeTumurtushaa2014ObsGB}
S.~Koh, B. -H.~Lee, W.~Lee, and G.~Tumurtushaa,
Phys.\ Rev.\ D \textbf{90}, 063527 (2014),
arXiv:1404.6096 [gr-qc].

\bibitem{vandeBruckDimopoulosLongden2016ReheatingGB}
C.~van de Bruck, K.~Dimopoulos, and C.~Longden,
Phys.\ Rev.\ D \textbf{94}, 023506 (2016),
arXiv:1605.06350 [astro-ph.CO].

\bibitem{BhattacharjeeMaityMukherjee2017ReheatingUnitarityPlanck}
S.~Bhattacharjee, D.~Maity, and R.~Mukherjee,
Phys.\ Rev.\ D \textbf{95}, 023514 (2017),
arXiv:1606.00698 [hep-th].

\bibitem{KohLeeTumurtushaa2018ReheatingGW}
S.~Koh, B. -H.~Lee, and G.~Tumurtushaa,
Phys.\ Rev.\ D \textbf{98}, 103511 (2018),
arXiv:1807.04424 [gr-qc].

\bibitem{GrandaJimenezTorres2021KineticAndGB}
L.~N.~Granda and D.~F.~Jimenez,
Eur.\ Phys.\ J.\ C \textbf{81}, 10 (2021),
\href{https://doi.org/10.1140/epjc/s10052-020-08789-9}{doi:10.1140/epjc/s10052-020-08789-9}.

\bibitem{PhamNguyenDo2021kGB}
T.~M.~Pham, D.~H.~Nguyen, and T.~Q.~Do,
arXiv:2107.05926 [gr-qc].

\bibitem{NozariShafizadeh2017BlueSpectrum}
K.~Nozari and S.~Shafizadeh,
Int.\ J.\ Mod.\ Phys.\ D \textbf{26}, 1750016 (2017),
arXiv:1712.09530 [gr-qc].

\bibitem{KalloshLindeRoest2013}
R.~Kallosh, A.~Linde, and D.~Roest,
JHEP \textbf{11}, 198 (2013)
[arXiv:1311.0472 [hep-th]]
\href{\doi/10.1007/JHEP11(2013)198}{doi:10.1007/JHEP11(2013)198}.












\end{thebibliography}
\end{document}